\documentclass[aps,prd,preprint]{emulateapj}
\usepackage{amsmath}
\usepackage{subfigure}
\usepackage{graphicx}
\usepackage{epsfig}
\usepackage{color}
\usepackage{units}
\usepackage{multirow}
\usepackage{mathptmx}

\newcommand{\V}[1]{ \mathbf #1 }

\begin{document}

\title{Ab initio Simulations of a Supernova Driven Galactic Dynamo in an
  Isolated Disk Galaxy}

\author{Iryna Butsky$^{1,2}$, Jonathan Zrake$^2$, Ji-hoon Kim$^2$, Hung-I Yang$^2$, Tom Abel$^2$}

\affil{Astronomy Department, University of Washington, Seattle, WA 98195, USA}
\affil{Kavli Institute for Particle Astrophysics and Cosmology, Stanford
  University, Menlo Park, CA 94025, USA}


\begin{abstract}
  We study the magnetic field evolution of an isolated spiral galaxy, using
  isolated Milky Way-mass galaxy formation simulations and a novel prescription for
  magnetohydrodynamic (MHD) supernova feedback.  Our main result is that a galactic
  dynamo can be seeded and driven by supernova explosions, resulting in magnetic
  fields whose strength and morphology is consistent with observations. In our
  model, supernovae supply thermal energy, and a low level magnetic field along
  with their ejecta. The thermal expansion drives turbulence, which serves a
  dual role by efficiently mixing the magnetic field into the interstellar
  medium, and amplifying it by means of turbulent dynamo. The computational
  prescription for MHD supernova feedback has been implemented within the
  publicly available \emph{ENZO} code, and is fully described in this
  paper. This improves upon \emph{ENZO}'s existing modules for hydrodynamic
  feedback from stars and active galaxies. We find that the field
  attains $\unit[]{\mu G}$-levels over $\unit[]{G yr}$-time scales throughout
  the disk. The field also develops large-scale structure, which appears to be
  correlated with the disk's spiral arm density structure. We find that seeding
  of the galactic dynamo by supernova ejecta predicts a persistent correlation
  between gas metallicity and magnetic field strength. We also generate all-sky
  maps of the Faraday rotation measure from the simulation-predicted magnetic
  field, and present a direct comparison with observations. 
\end{abstract}

\maketitle

\section{Introduction}

Magnetic fields exist throughout all types of galaxies. In spiral galaxies like
our own, the fields typically attain several $\unit[]{\mu G}$ in strength, and
exhibit long-range structure throughout the disk \citep[see e.g.][for recent
  reviews]{Beck2009, Beck2015}. The processes by which these fields have
originated, and how they influence their host galaxy's evolution remains the
subject of intense observational and theoretical research. On the one hand,
understanding the build-up of magnetic field by the galactic flow \citep[the
  dynamo problem, see e.g.][]{Moffatt1978, Parker1979} poses significant
analytical and computational challenges. On the other, a galaxy's magnetic field
influences its pressure and flow profiles \citep{Elstner2014}, rate of star
formation \citep{VanLoo2014, Federrath2012, Federrath2013, Federrath2015}, and
cosmic ray transport processes \citep[e.g.][]{Strong2007}. The co-evolution of
galaxies and their magnetic fields is thus a complex and deeply significant
problem.

The history of our galaxy's magnetic field certainly includes a type of
 dynamo process, which has enhanced a pre-existing ``seed'' magnetic field and
now maintains it at near-equipartition levels. It has long been appreciated that
if the dynamo's exponential folding time is the galactic rotation period, then
it must have been seeded around the time of the Milky Way's formation at a level
not less than $\sim \unit[10^{-20}]{G}$ \citep{Rees1987}. This has motivated
extensive consideration of early universe processes that might have magnetized
the primordial plasma from which the galaxy formed. However, a more modern view
is that seeding by means of astrophysical processes, such as battery mechanisms,
kinetic instabilities, or stellar feedback, would be of greater consequence than
a primordial seed field \citep[e.g.][]{Blackman1998, Rees2006}. It is also
appreciated now that dynamo action is expected to grow the magnetic field at the
turbulent diffusion rate, which is significantly faster than the orbital shear
\citep{Kulsrud1992}.

In this paper we address the question of how the magnetic field in a disk galaxy
like the Milky Way would have evolved under the influence of stellar feedback
alone. Within this narrative, the first significant magnetic fields would have
been generated by turbulent convection within the interiors of early stars, and
then dispersed by their winds or supernova explosions. Such feedback would
inject magnetized plasma into the interstellar medium (ISM), while
simultaneously driving turbulence that acts to disperse and amplify the
field. We are thus investigating whether stars might have served as
the seed for, and also the engine of galactic dynamo action. Our main result
will be that this scenario offers a compelling resolution to the mystery of
galactic magnetism, as it turns out to predict a magnetic field whose strength
and morphology are consistent with what is known about the Milky Way's magnetic
field.

Our results are based on simulations of an initially \emph{unmagnetized}
galactic disk, coupled with a realistic prescription for star formation and
magnetohydrodynamic (MHD) supernova feedback. In this prescription, supernova
events occur in proportion to star formation activity, and provide localized
injections of thermal energy, metal-rich material, and a low level of toroidal
magnetic field. This is in contrast to a number of earlier studies in which
supernovae were assumed to function strictly as sources of energetic cosmic rays
\citep{Hanasz2009, Siejkowski2014, Kulpa-Dybe2011, Kulpa-Dybe2015}. In
this study, we will neglect the influence of cosmic rays, and focus on the role
of turbulence driven by supernovae through their thermal feedback. The fact that
we ignore the cosmic rays, and yet recover results that are broadly consistent
with earlier studies, suggests that galactic magnetism is a robust phenomenon,
arising in response to several processes independently.

This computational approach will allow us to flesh out a number of details that
have yet to be addressed within the supernova-driven galactic dynamo
scenario. First, we wish to understand whether seed fields that are supplied
locally, in small volumes around supernovae, can be mixed efficiently throughout
the disk. We anticipate that if this mixing is efficient, \footnote{Simulations
  in a more idealized setting do suggest this is the case, see \cite{Cho2012}.}
then the outcome of our simulations may not differ dramatically from earlier
ones in which a low-level seed field was imposed uniformly on the initial data
\citep{Wang2009, Rieder2016}. We will also examine how seeding of magnetic
fields by metal-rich supernova ejecta might induce a persistent correlation
between metallicity and magnetism in the ISM. This is a non-trivial question
because metals and magnetic flux experience distinct types of advective
transport (scalar versus vector). Finally, there is the question of whether a
supernova-driven dynamo leads to the development of large-scale magnetic
structure. All of these questions will be addressed in the present work.

There have been a good number of studies focused on other aspects of the
supernova-driven galactic dynamo. It has been simulated in a localized setting,
with and without the effects of orbital shear \citep{Gressel2008,
  Gressel2008a}. Global simulations were carried out by \citet{Wang2009}, using
\emph{ENZO} to study the formation and evolution of an isolated galaxy, onto
which a seed field of $\unit[10^{-9}]{G}$ was imposed as part of the initial
data. While that paper provides valuable insight on galactic magnetic field
amplification, it does not include molecular cooling, star formation, or
supernova feedback. \citet{Xu:2009} used magnetic AGN feedback in 
galaxy clusters to show that turbulence in the ICM can amplify
magnetic fields to observed values. \citet{Schleicher:2010} investigated 
magnetic field growth via the small-scale dynamo for Kolmogorov and Burgers-type turbulence during the formation 
of the first stars and galaxies. \citet{Dubois:2010aa} introduced a more efficient method of
including thermal supernova feedback in cosmological simulations using
\emph{RAMSES} and showed that dwarf galaxies seeded by magnetized supernova
bubbles can attain magnetic fields of $\mu$G strength in 1 Gyr. \citep{Teyssier:2002} and studied its effect on the evolution of
different types of galaxies. \citet{Beck:2012aa} used the SPMHD code \emph{GADGET}
\citep{Springel:2001} to show that a primordial seed field can be amplified to equipartition
strength in Milky-Way type galactic halos during virialization with the help of a small
scale dynamo driven by supernova feedback. \citet{Pakmor:2013aa} observed exponential magnetic
field amplification using an improved MHD implementation \emph{AREPO}
\citep{Pakmor:2011}. However, instead of explicitly including thermal supernova
feedback, they incorporated feedback with a modified equation of state.
\citet{Rieder:2015} use \emph{RAMSES} to build upon the work of
\citet{Dubois:2010aa} and provide an in-depth study of the galactic
amplification of a primordial magnetic field in the presence of thermal
supernova feedback. Though they demonstrate that the turbulence created by
supernova feedback drives the small-scale dynamo that amplifies the original
magnetic field, they do not include magnetized supernova feedback.

\citet{Beck:2013aa} incorporated magnetic seeding by supernova remnants in
cosmological zoom-in simulations using \emph{GADGET}. 
In their model, they inject a dipolar magnetic field in
the range of $\unit[10^{-5}-10^{-3}]{G}$ into a volume of linear dimension 5 pc
surrounding the supernova, which ultimately expands into a 25 pc radius
bubble. The initial magnetic field is amplified to $\unit[]{\mu G}$ strengths
through turbulent dynamo resulting from gravitational collapse, supernova
feedback, and galactic mergers. They point out that even if a primordial
magnetic seed field were present, it would have been washed out by the stronger
fields supplied by supernovae. The goals of \citet{Beck:2013aa} differ from our
own in that their analysis focuses on a protogalactic halo, before it develops
into a disk. Here we will study an isolated galactic disk, and neglect its
merger history. We also model our injected magnetic fields with a toroidal
geometry instead of poloidal, although we do not anticipate this difference to
be consequential. Finally, we note that the inclusion of cosmic rays in
simulations of galaxy formation is quickly becoming feasible; algorithms that
afford a self-consistent treatment of the cosmic ray fluid within cosmological
MHD frameworks have been described in \citet{Salem2014, Pfrommer2016,
  Pakmor2016, Pakmor2016a}. Our choice to explore the role of thermal, as
opposed to cosmic ray feedback processes, is based on a wish to understand the
minimal conditions from which galactic magnetic fields are expected to arise.

Our paper is organized as follows. In \S \ref{sec:methods}, we describe our
prescription for stellar formation and feedback, devoting particular attention to
the numerical details of the magnetic field source term associated with the MHD
supernova feedback. In \S \ref{sec:results}, we present the results of our
simulations, including a resolution study, images depicting the evolved magnetic
field morphology, time series of the galactic magnetic energy budget, the power
spectral distribution of magnetic energy, and correlation of magnetic field
strength with gas density and metallicity. In \S \ref{sec:rotation-measures} we
present all-sky maps of the Faraday rotation measure generated from simulation
data. Our findings are summarized in \S \ref{sec:conclusion}. There, we also
draw comparisons with other numerical studies, speculate as to the mechanism by
which our simulated galaxy develops its long-range magnetic field, and offer
some reflection on the appearance of the earliest galactic magnetic fields.

\begin{figure*}[t]
    \centering
    \leavevmode\epsfxsize=18cm\epsfbox{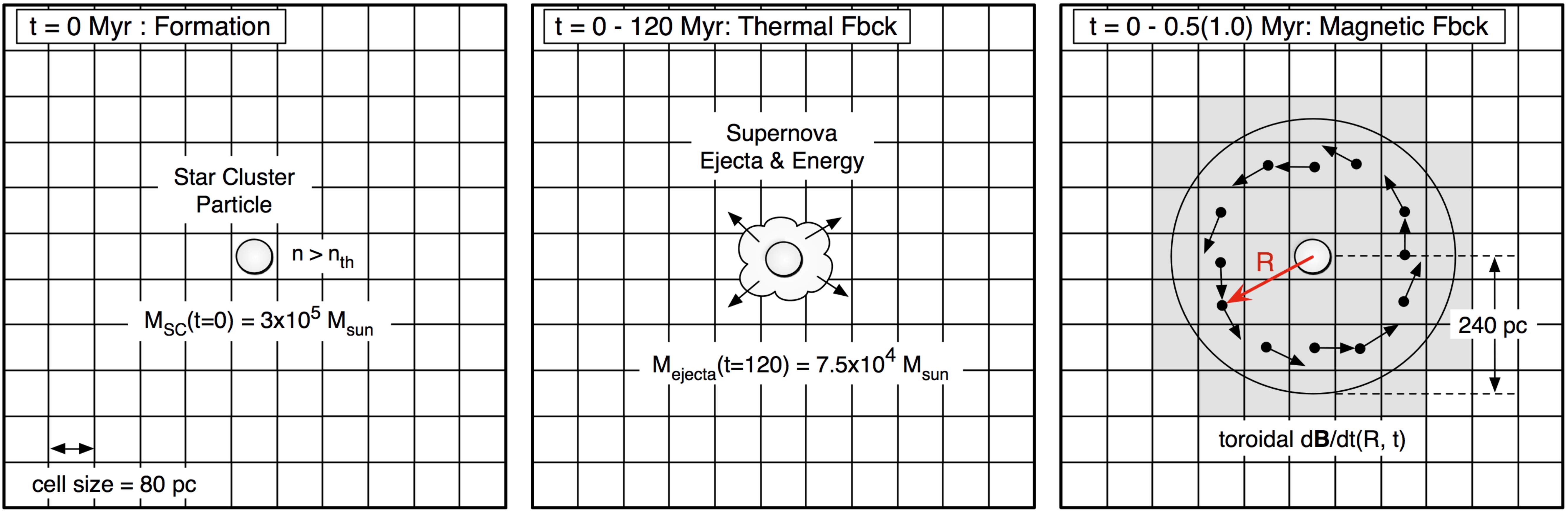}
    \caption{Two-dimensional schematic overview of the life cycle of a star
      cluster particle and two channels of its feedback.  {\it Left:} Star
      cluster particle formation for an 80 pc resolution run (g80; see \S
      \ref{sec:formation}).  {\it Middle:} Thermal feedback (see \S
      \ref{sec:thermal_fbck}).  Thermal energy by Type II supernova explosion is
      injected into the gas cell in which a star cluster particle of age less
      than 120 Myr resides .  {\it Right:} Magnetic feedback (see \S
      \ref{sec:magnetic_fbck}).  Toroidal magnetic fields are seeded within
      three finest cells from a star cluster particle.
\label{fig:BTF}
}
\end{figure*}

\section{Numerical Methods}
\label{sec:methods}

\subsection{Refinement, Hydrodynamics, and Radiative Cooling}

We conduct this study by simulating isolated disk galaxies at three different
resolutions using {\it ENZO}, a multi-physics cosmological magnetohydrodynamic
(MHD) code that uses adaptive mesh refinement technology
\citep[AMR;][]{Collins:2010, Bryan:2014aa}.  {\it ENZO} includes a wide range of
relevant physics previously considered in galaxy-scale simulations
\citep[e.g.][]{Kim:2011, Kim:2013a, Kim:2013b} some of which are described in
detail below.

The AMR function allows us to achieve higher resolutions more efficiently by
only fully resolving areas of interest, designated by baryon and particle
overdensities.  Effectively, the entirety of our galactic disks were resolved at
the highest level, while the outer edges of the simulation were able to remain
less resolved to conserve computing time.  The mass thresholds above which a
cell is refined by factors of two in each axis are $M_{\rm ref, gas} =
2.15\times10^4 M_{\odot}$ and $M_{\rm ref, part} = 1.72 \times 10^6 M_{\odot}$
for gas and particles, respectively.  In addition, the local Jeans length is
always resolved by at least four cells to avoid artificial fragmentation
\citep{Truelove:1997}.  We use a variety of grid configurations, ranging from a
$64^3$ initial grid for a simulation box of 1.31 Mpc with 6 levels of
refinement, to a $128^3$ initial grid with 7 levels of refinement.  A
combination of these grid conditions allows us to simulate a galaxy (described
in Table 1) with three different minimum grid sizes of 320, 160, and 80 pc.

\begin{table*}
  \begin{minipage}{180mm}
    \begin{center}
      \caption{Simulation Parameters}
      \begin{tabular}{l || c | c | ccccc}
        \hline \hline Run ID & Minimum grid size (pc) & Disk initial metallicity
        ($Z_{\odot}$) & \multicolumn{5}{c}{Supernova magnetic feedback
          parameters \footnote{ \scriptsize $L$ and $\tau$ are the
            characteristic distance and time scale of a supernova magnetic
            feedback event, respectively.  $r_{\rm cutoff}$ and $t_{\rm cutoff}$
            are the cutoff distance and time for magnetic feedback influence,
            respectively.  $E_{\rm tot, \,B} = \sigma E_{\rm tot}$ corresponds
            to the total magnetic energy injected over all space and time per
            each supernova feedback event.  For detailed explanation of these
            magnetic feedback-related parameters, see \S
            \ref{sec:magnetic_fbck}.}} \\ \cline{4-8} & & & $L$ (pc) & $r_{\rm
          cutoff}$ (pc) & $\tau$ (Myr) & $t_{\rm cutoff}$ (Myr) & $E_{\rm tot,
          \,B}$ (ergs) \\ \hline g320 & 320 & 1.0 & 262 & 786 & 0.2 & 1 &
        $3.3\times 10^{48}$ \\ g160 & 160 & 1.0 & 131 & 393 & 0.1 & 0.5 &
        $3.3\times 10^{48}$ \\ g160LM & 160 & 0.001 & 131 & 393 & 0.1 & 0.5 &
        $3.3\times 10^{48}$ \\ g160LR & 160 & 1.0 & 262 & 786 & 0.1 & 0.5 &
        $3.3\times 10^{48}$ \\ g80 & 80 & 1.0 & 79 & 236 & 0.025 & 0.1 &
        $3.3\times 10^{48}$ \\ g80LR & 80 & 1.0 & 262 & 786 & 0.1 & 0.5 &
        $3.3\times 10^{48}$ \\ \hline
      \end{tabular}
    \end{center}
  \end{minipage}
  \label{tab:parameters}
\end{table*}

In order to solve the fluid conservation equations, we use a hyperbolic
divergence cleaning approach with the MHD method that is described in
\citet{Dedner:2002} and is extensively tested in {\it ENZO} by \citet{Wang2009}
with various set-ups including an idealized disk formation simulation.  In the
work presented here, time integration is carried out by the total variation
diminishing (TVD) 2nd order Runge-Kutta (RK) scheme \citep{Shu:1988}.  Spatial
reconstruction employs the piecewise linear method \citep[PLM;][]{VanLeer:1977},
and the flux at cell interfaces is computed with the local Lax-Friedrichs
Riemann solver \citep[LLF;][]{Kurganov:2000}.  A maximum 30\% of the
Courant-Friedrichs-Lewy (CFL) timestep is used to advance any fluid element in
the simulation (i.e. hydrodynamic CFL safety number of 0.3).  For a detailed
description and testing of the MHD machinery we adopt, including the Dedner
formulation of MHD equations in {\it ENZO}, we refer interested readers to
\citet{Wang:2008}, \citet{Wang2009}, and references therein.

In addition, the gas in the ISM of our galaxy cools radiatively.  Our
equilibrium cooling follows pre-computed tabulated cooling rates from the
photoionization code {\sc Cloudy} \citep{Ferland:2013} provided via the {\sc
  Grackle} chemistry and cooling library \citep{Bryan:2014aa, Kim:2014aa}, that
is plugged in to {\it ENZO}.  The pre-computed look-up table includes metal
cooling rates for solar abundances as a function of gas density and temperature.
These metal cooling rates will be scaled linearly with the metallicity that is
being traced as a passive scalar field throughout our simulations.  Uniform
photoelectric heating of $8.5 \times 10^{-26}\,\, {\rm erg \,s^{-1} cm^{-3}}$ is
also considered without self-shielding \citep{Tasker:2008}.  We however do not
include the photoionization heating by the metagalactic ultraviolet background
radiation from quasars and galaxies, leaving it for future exploration of the
galaxy formation parameter space.

\subsection{Star Cluster Particle Formation}
\label{sec:formation}

The prescription for star formation and supernova feedback follows the algorithm
first described in \citet{Cen:1992} with minor modifications.  The finest cell
(at the maximum level of refinement) of size $\Delta\, x$ and gas density
$\rho_{\rm gas}$ produces a star cluster particle of initial mass $M_{\rm
  SC}^{\rm init} = 0.1\, \rho_{\rm gas} \Delta \,x^3$ if all of the conditions
below are met:
\begin{itemize}
\item[{\it (a)}] the gas density is over $\unit[3.0 \times 10^{-26}]{g \,
  cm^{-3}}$,
\item[{\it (b)}] the gas flow is converging (velocity field has negative
  divergence),
\item[{\it (c)}] the cooling time $t_{\rm cool}$ is shorter than the gas
  dynamical time $t_{\rm dyn}$ of the cell,
\item[{\it (d)}] the cell mass is larger than the local Jeans mass (Jeans
  unstable), and
\item[{\it (e)}] the cell has enough mass to create a particle of at least
  $M_{\rm thres} = 3.0 \times 10^5 \, M_{\odot}$.
\end{itemize}
There are some notable differences between these conditions for star formation,
and those adopted by other studies that also utilized \emph{ENZO}
\citep[e.g.][]{Tasker:2008}. First, the prefactor $\Delta t / t_{\rm dyn}$ in
Equation 1 of \citet{Tasker:2008} is removed in order not to leave any
unresolved mass behind when a star cluster particle is formed. Second, we do not
impose ``stochastic star formation''. And third, once created, a star cluster
particle supplies thermal feedback (detailed below) into its surroundings for a
fixed period of $\unit[120]{Myr}$ (rather than $12 \, t_{\rm dyn}$).

\subsection{Supernova Feedback: Thermal Energy Injection}
\label{sec:thermal_fbck}

Of the initial mass of a star cluster particle deposited, $M_{\rm SC}^{\rm
  init}$, 75\% is eventually locked up in the particle after 12 $t_{\rm dyn}$.
The {\it actual} stellar mass, $M_{*}$, formed inside a star cluster particle is
described as a function of particle age $t$ by
\begin{align}
  M_{*}(t) &= 0.75 \, M_{\rm SC}^{\rm init} \int_0^{\bar t} {\bar t}\,'
  e^{-{\bar t}\,'}\,d{\bar t}\,' \\ &= 0.75 \, M_{\rm SC}^{\rm init} \left[ 1-
    (1+ {\bar t}) e^{-\bar t} \right] \, ,
  \label{eq:SF}
\end{align}
where ${\bar t} = t/t_{\rm dyn}$.

Note that the {\it actual} stellar mass formation peaks at one dynamical time of
the birthplace gas cell, $t_{\rm dyn}$.  The rest of the particle mass,
0.25\,$M_{\rm MC}^{\rm init}$, gradually returns to the cell in which the
particle resides, along with $10^{-5}$ of the rest mass energy of $M_*$ (see
Figure \ref{fig:BTF}).  This models $10^{51}$ ergs of thermal energy per every
42 $M_{\odot}$ of {\it actual} stellar mass formed.  It is equivalent of Type II
supernova explosion's energy input and is enough to replenish the energy loss in
ISM due to radiative cooling.  Therefore, for a star cluster particle of initial
mass $M_{\rm SC}^{\rm init} = M_{\rm thres} = 3 \times 10^5 \,M_{\odot}$ which
eventually produces $M_* = 2.25 \times 10^5 \,M_{\odot}$, its supernova feedback
event injects a total thermal energy of $E_{\rm tot, \,th} = 5.4 \times 10^{54}$
ergs.  As discussed in \S \ref{sec:magnetic_fbck}, we will consider the case
where the injected thermal energy dominates the total supernova feedback energy
budget (i.e. $E_{\rm tot, \,B} \ll E_{\rm tot, \,th} $ thus $E_{\rm tot} \simeq
E_{\rm tot, \,th} $).  The ejecta mass blown back into the gas phase of ISM is
recycled in the next generation of star formation.  2\% of the ejecta mass is
considered to be metals.

\subsection{Supernova Feedback: Magnetic Field Injection}
\label{sec:magnetic_fbck}

Balancing numerical resolution and the computational feasibility of a simulation
is one of the major challenges in large numerical calculations.  For this
reason, in this work we aim to build a prescription for the magnetic field
injection by supernova remnants with a simple geometry that could accurately be
resolved by as few grid cells as possible.  We propose that the best
approximation is to source a toroidal loop of magnetic field around a star
cluster particle (of mass $\sim 3 \times 10^5 \,M_{\odot}$) that generates a
supernova feedback event (see Figure \ref{fig:BTF} and \S
\ref{sec:thermal_fbck}).

The spatial and temporal evolution of the injected magnetic energy is chosen to
be
\begin{equation} \label{eqn:Ub-dot-source}
  \dot{U}_{B,\, {\rm source}} = \tau^{-1} \frac{B_0^2}{4\pi} \frac{R}{L}
  e^{-r^2/L^2} e^{-t/\tau} (1-e^{-t/\tau})
\end{equation}
where $t$ is the star cluster particle age, $R$ is the cylindrical radius, $r$
is the spherical radius, and $L$ and $\tau$ are the characteristic distance and
time scale of a supernova magnetic feedback event, respectively. The time
profile in Equation \ref{eqn:Ub-dot-source} increases from zero at $t=0$ to its
characteristic value $\sim B_0^2 / \tau$ around $t \sim \tau$, and then declines
exponentially to zero with decay constant $\tau$.

By integrating the magnetic energy of the source field over all space and time,
we acquire the expression for the normalization factor $B_0$ as
\begin{equation} \label{eqn:B0-val}
  B_0^2 = \frac{64}{3} \frac{\sigma E_{\rm tot}} {V_{\mathrm{SN}}} \, ,
\end{equation}
where $E_{\rm tot} $ is the total energy injected by a supernova feedback event,
$\sigma $ is the proportion of the magnetic energy of the supernova to the total
energy of the supernova, and $V_{\mathrm{SN}} = \frac{4}{3}\pi L^3$ is the
volume of the supernova magnetic feedback event. Equations
\ref{eqn:Ub-dot-source} and \ref{eqn:B0-val}, combined with the relation $\dot
U_{B, \, {\rm source}} = \mathbf{B} \cdot \dot{\mathbf{B}} / 4 \pi$ and our
assumption that injected magnetic field is toroidal (along the randomly oriented local azimuthal
unit vector $\hat{\mathbf{e}}_\phi$ for each event), yield an expression for the
source term of magnetic field,
\begin{equation} \nonumber
  \dot{\mathbf{B}}_{\rm source} = \tau^{-1} B_0 \left(\frac{R}{L}\right)^{1/2}
  e^{- r^2 / 2L^2} e^{-t / \tau} \, \hat{\mathbf{e}}_\phi \, .
\end{equation}

In the presented simulations, we consider the case where the injected thermal
energy dominates the total supernova feedback energy budget, that is, $E_{\rm
  tot, \,B} \ll E_{\rm tot, \,th} $ thus $E_{\rm tot} \simeq E_{\rm tot, \,th}
$.  For each supernova feedback event, we choose to inject a total magnetic
energy of $E_{\rm tot, \,B} = \sigma E_{\rm tot} =
\unit[3.3\times10^{48}]{erg}$, implying $\sigma = 6.1 \times 10^{-7}$ for a star
cluster particle of initial mass $3 \times 10^5 \,M_{\odot}$ (see \S
\ref{sec:thermal_fbck}). In addition, we set the characteristic distance scale
of the magnetic feedback to be comparable with the minimum grid size of each
run.  For our choices of characteristic scale parameters $L$ and $\tau$ for each
run, we refer the readers to Table 1.  Lastly, because of numerical
consideration, a cutoff distance and time are imposed when implementing the
aforementioned $\dot{U}_{B,\, {\rm source}}$ and $\dot{\mathbf{B}}_{\rm source}$
formulae. We fiducially choose to cut off the exponential functions at $r_{\rm
  cutoff} = 3L$ and at $t_{\rm cutoff} = 5\tau$. Since limited numerical
resolution requires that $L$ be much larger than the scale of an actual
supernova remnant, we choose $L$ to be as small as possible, while remaining
smoothly resolved on the grid. In practice, we find that numerical consistency
between the actual magnetic energy introduced, and the integral over space and
time of Equation \ref{eqn:Ub-dot-source}, is acceptable when the supernova event
occupies a subset of the computational domain as small as $3^3$ grid cells (see
Figure \ref{fig:BTF}). In our resolution comparison runs, g80LR and g160LR, the 
injection stencil matches that of the least-resolved run, g320.

\subsection{Initial Condition}
\label{sec:IC}

Our simulated galaxies are evolved from one of the isolated disk galaxy initial
conditions (ICs) identified by the AGORA Collaboration \citep{Kim:2014aa}, which
strives to compare a wide range of galaxy simulations by creating a standardized
set of ICs and feedback models.  While the details of the proposed AGORA ICs are
presented in the aforementioned paper, we summarize the aspect of the isolated
ICs that might be of interest to readers.  Note that by adopting isolated
galactic disk ICs, we choose to ignore the effect of major and minor mergers.
Here we already make an implicit assumption that such inter-galactic-scale
dynamical interactions are secondary in the buildup of the galactic magnetic
field.

The AGORA isolated IC was initially generated with the {\sc Makedisk} code and
distributed to the community as ASCII files in three resolution options.  For
our study we adopted the low-resolution version.  The disk has a total mass of
$M_{\rm d}=4.297\times 10^{10} \,\, M_{\odot}$, 80\% of which is in $10^5$
stellar particles, and the rest 20\% ($=f_{\rm gas}$) is in a gaseous disk that
follows an analytic exponential profile
\begin{equation*}
  \rho(r,\,z) = \rho_0 \ e^{-r/r_{\rm d}} \ e^{-|z|/z_{\rm d}}
\end{equation*}
with scale length $r_{\rm d} =$ 3.432 kpc, scale height $z_{\rm d} =$ 343.2 pc,
and $\rho_0= M_{\rm d}f_{\rm gas} / 4\pi r_{\rm d}^2 z_{\rm d}$.  The disk has
an initial temperature of $10^4$~K, and an initial metallicity of $Z_{\rm d,
  init} = 1\,Z_{\odot}$ for the g320, g160, g160LR, g80, g80LR runs, and $10^{-3}\, Z_{\odot}$
for the g160LM run (see Table 1).  The stellar bulge of $1.25\times 10^4$
particles follows the Hernquist density profile \citep{Hernquist:1990} with a
bulge to disk mass ratio of 0.1.  The dark matter halo has a mass of $M_{200} =
1.074\times10^{12} \,\,M_{\odot}$ in $10^5$ particles, and follows the
Navarro-Frenk-White profile \citep[NFW;][]{Navarro:1997} with concentration
parameter $c=10$ and spin parameter $\lambda=0.04$.  This system is embedded in
hot ($10^6$~K) gaseous medium, uniform in density across the entire simulation
box of $(1.31 \,\,{\rm Mpc})^3$.  However the sum of this {\it halo} gas mass is
only equal to the disk stellar mass, with zero velocity and negligible
metallicity ($10^{-6}\,Z_{\odot}$), so it has negligible effect in the disk's
evolution.  This IC models typical structural properties and gas fraction that
are characteristic of Milky Way-like galaxies at $z \sim 1$.

\section{Results}
\label{sec:results}

Here we describe the results of a family of simulated disk galaxies, evolved
using the MHD supernova feedback model and initial conditions described in \S
\ref{sec:methods}. Our family of models is summarized in Table
\ref{tab:parameters}. It is parameterized around the mesh spacing (in parsecs)
of the finest grid blocks (those on the galactic plane). For example, in the run
g160, the smallest cells are 160 parsecs in size. The scale $L$ over which
supernova feedback is active is chosen to be smaller with increasing resolution,
such that $L$ is roughly 3 grid cells across in each model. The composition
of the supernova ejecta is given a metallicity of $Z_\odot$, with the exception
of an additional run, g160LM, for which the metallicity was chosen to be $10^{-3}
Z_\odot$. This choice allows us to rule out the possible influence of chemistry
dependent cooling on the magnetic field evolution.

\subsection{Magnetic Field Morphology}
\label{sec:morphology}

\begin{figure}[t]
    \centering
    \leavevmode\epsfxsize=9.3cm\epsfbox{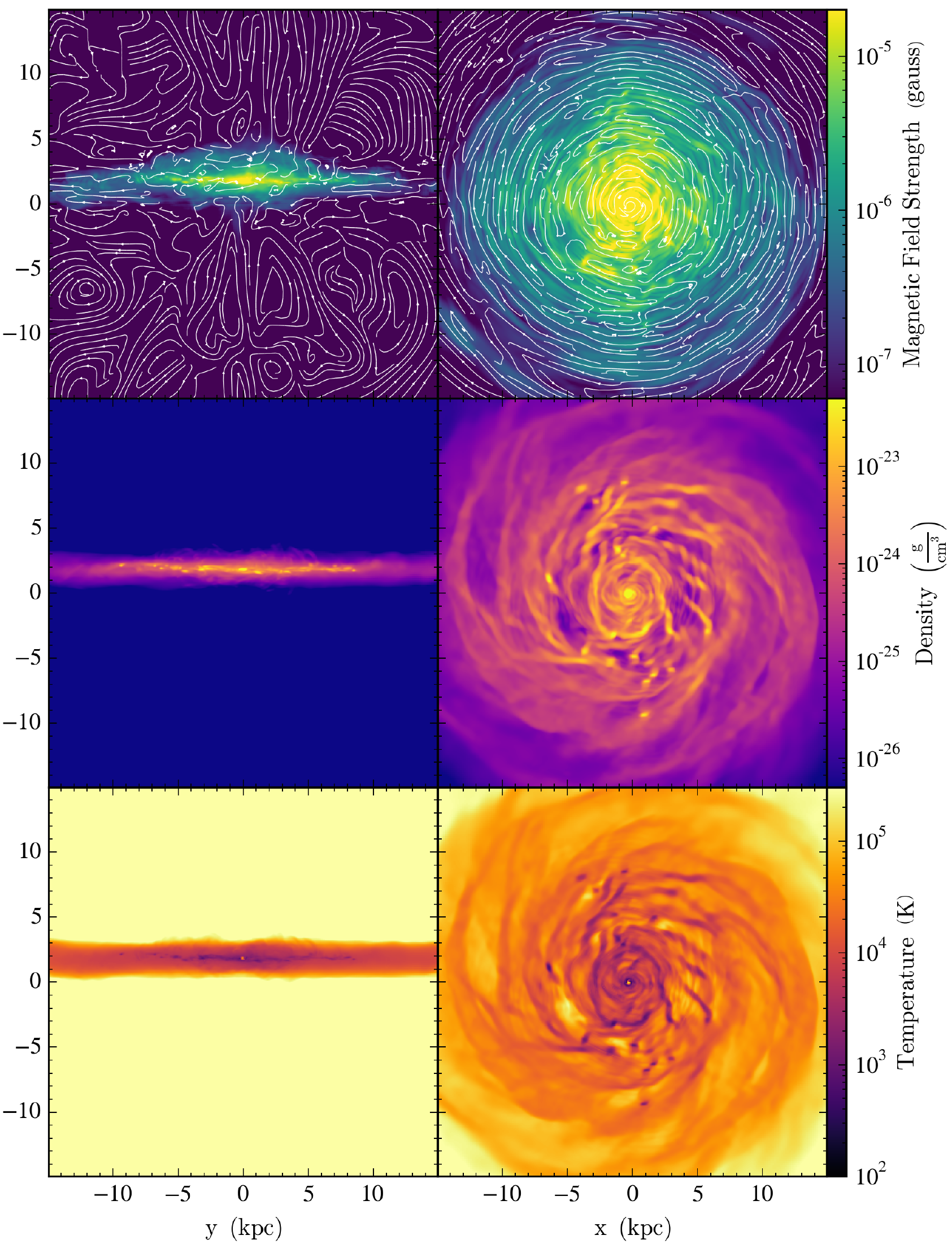}
    \caption{\footnotesize The edge-on {\it (left)} and face-on {\it (right)}
      projections of the magnetic field strength {\it (top)}, density {\it
        (center)}, and temperature {\it (bottom)} of galaxy g80LR at t = 2.1
      Gyr in 30 kpc boxes. Magnetic field streamlines are plotted in black over
      the image of the magnetic field strength.
\label{fig:projections}
}
\end{figure}

The magnetic field in our disk galaxy simulations is illustrated in Figure
\ref{fig:projections}. Shown there are density-weighted projections of magnetic
and thermodynamic variables, taken 2.1 Gyr into the simulation at the highest
resolution (model g80LR), with the disk shown edge-on in the left column, and
top-down in the right. At this time, saturation of the magnetic energy
throughout the disk is complete (see \S \ref{sec:time-series}). Images are
zoomed into the central $\unit[35]{kpc}$ of the simulation. The top row shows
that the magnetic field strength attains levels of several $\unit[]{\mu G}$ near
the disk mid-plane, and within roughly $\unit[10]{kpc}$ of the galactic
center. In that image, we have also plotted streamlines of the magnetic field
(projected onto the disk plane) to provide an impression of its overall
geometry. From the top-down image, one sees that the field throughout the disk
is predominantly azimuthal, with the radial field component $B_R$ being
significantly lower than $B_\phi$. We believe this indicates that differential
rotation of the disk is active at sustaining the field strength around its
saturated value. Prevalence of the azimuthal field is consistent with
observations e.g. IC 342, where the magnetic pitch angle is inferred to be
relatively small \citep{Sokoloff1992}. We also note that the azimuthal field in
our simulations changes sign at several radii. Those reversals imply the
presence of magnetic neutral lines, which have been observed in M81
\citep{Krause1989}. The density profile indicates higher-density spiral arms,
which correspond to areas of lower temperature and higher magnetic field
strength.

The left column of Figure \ref{fig:projections} illustrates the field morphology
projected onto the meridional plane. The relative isotropy of meridional
streamlines indicates that vertical and radial magnetic field components are
comparable to one another ($B_z \sim B_R$). It is also evident from the relative
smoothness of streamlines that the field is tangled at a smaller scale in the
vicinity of the disk mid-plane than in the halo. Finally, we see that the halo
magnetic field is considerably weaker than the field around the mid-plane. These
observations are consistent with the view that agitation of the gas by
supernovae, together with differential rotation of the disk, cooperatively
maintain the short wavelength, equipartition-level magnetic field.

In contrast, the field in the halo is seen to be longer wavelength, and
sub-equipartition. The relatively large coherency scale could be due to inverse
transfer effects known to accompany turbulent relaxation \citep{Zrake2014,
  Brandenburg2015, Zrake2016}, or to mild turbulent driving supplied by the
supernova outflows emerging from the disk. We favor the latter explanation
because the halo magnetic field strength is sub-equipartition with respect to
the kinetic energy density of turbulence, in other words motions driven
externally are super-Alfv\'{e}nic, so relaxation effects are not expected to be
dominant here. The relatively large coherence scale of the halo magnetic fields
is thus thought to arise from the presence of turbulent motions at a similar
scale, which are in turn driven by the supernova outflows.

It is noteworthy that well-ordered poloidal magnetic field in galactic halos
have been reported by e.g. \cite{Reuter1994}, based on analysis of the
$\unit[]{cm}$-wavelength synchrotron rotation measures of M82, and also by
\cite{Dahlem1997}, based on soft X-ray emission from high latitudes in NGC
4666. In both cases, the authors surmised the field regularity was due to
inertial effects of a galactic wind. While that may be the case, the halo
magnetic field in our simulated galaxies develops large-scale coherency even
though no wind is driven from the galactic disk.

\subsection{Magnetic Field Growth}
\label{sec:time-series}

\begin{figure}[t]
  \centering
  \leavevmode\epsfxsize=8.6cm\epsfbox{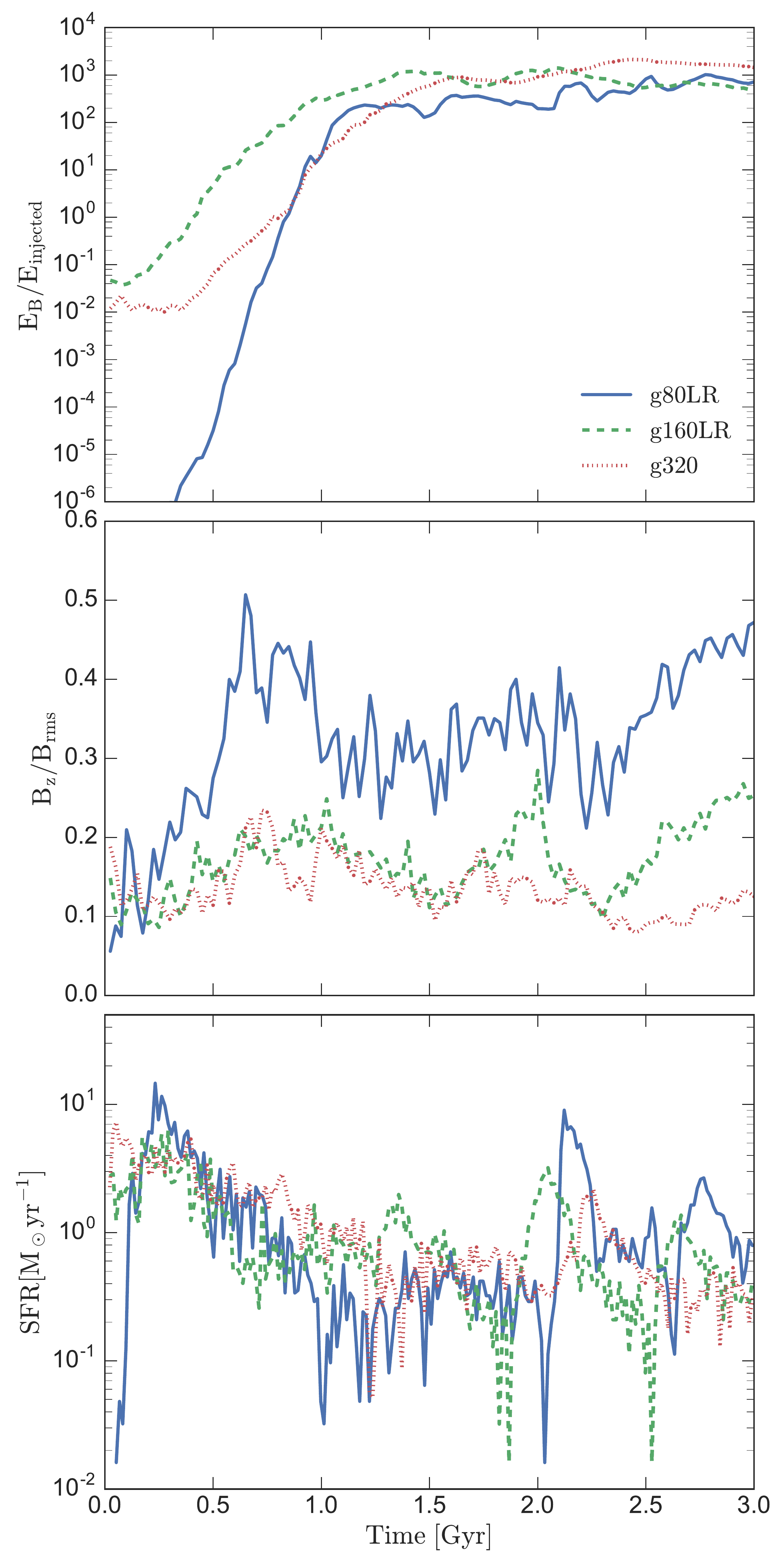}
  \caption{\footnotesize \emph{Top panel}: The ratio of total magnetic energy
    of 40 kpc around the center of the galaxy to the total injected magnetic energy 
    as a function of time.  \emph{Middle
      panel}: Proportion of the poloidal component of the magnetic field to the
    root-mean-square magnetic field value as a function of time. \emph{Bottom
      panel}: The star formation rate as a function of time.}
  \label{fig:time_series}
\end{figure}

\begin{figure}[t]
  \centering
  \leavevmode\epsfxsize=8.6cm\epsfbox{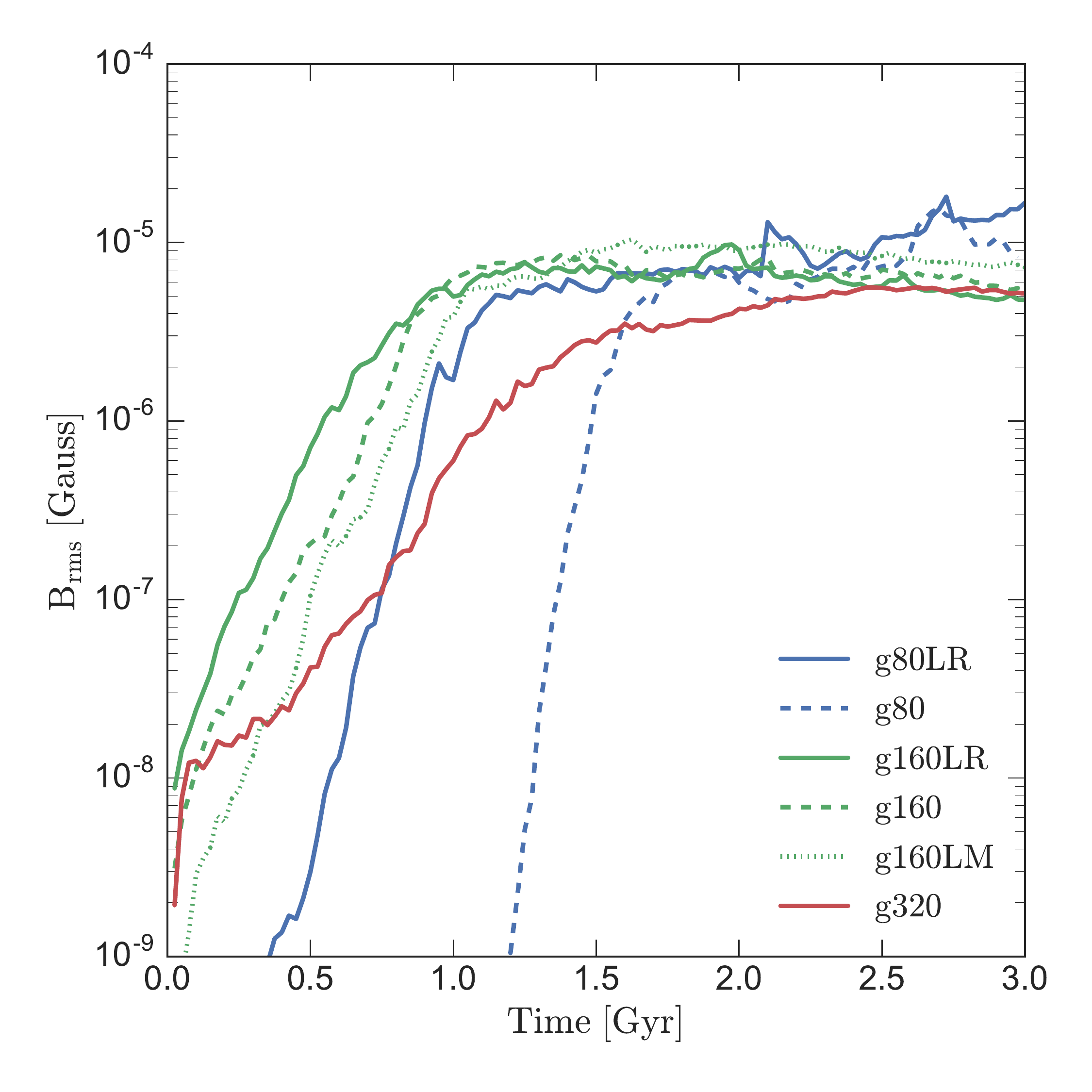}
  \caption{\footnotesize The density weighted root-mean-square magnetic field strength as a function 
of time. }
  \label{fig:B_rms}
\end{figure}

Our simulated disk galaxies are initialized in a completely unmagnetized
state. The appearance of magnetic fields accompanies the first supernova
explosions. Those fields are introduced at a very low level (only one part in
$\sim 10^{8}$ of the supernova energy is in magnetic form). From there,
enhancement of the magnetic field can only occur through some sort of dynamo
action, which we define broadly as any conversion of kinetic into magnetic
energy.

In the first phase of the magnetic field evolution, lasting roughly the first
billion years, each of the galaxy models undergoes exponential growth of their
total magnetic energy. In the top panel of Figure \ref{fig:time_series}, we show
the ratio total magnetic energy inside a $\unit[40]{kpc}$ sphere about the center of
the domain to the total injected magnetic energy from supernova events, 
as a function of time. We note that in the first 500
million years, the total magnetic energy injected by the supernovae is higher
than the total magnetic energy of the galaxy. This reflects that fact that
magnetic energy may be reduced by adiabatic expansion or lost due to numerical
dissipation, while net growth of the energy only occurs if dynamo action
overwhelms those processes. We speculate that very early in the simulation,
enhancement is weak because only the gas immediately nearby the supernova
feedback sites is magnetized. As the magnetized gas is dispersed by turbulent
mixing, amplification can occur rather uniformly throughout the disk. We believe
that net enhancement (exponential growth) of magnetic energy commences once this
initial mixing stage is complete. The exponential growth of the magnetic field
does not follow a single straight-line (in log space) path, but can rather be
divided into several stages of varying exponential growth rates. Model g320
reaches saturation at 1.5 Gyr, while the other three models reach saturation
after only 1 Gyr.

We believe that exponential growth of the magnetic energy is best explained by
operation of a small-scale turbulent dynamo. This process (also known as
Kraichnan-Kazenstev dynamo \citep{Kraichnan1968, Kazantsev1968}, see also
\cite{Brandenburg:2005, Tobias2011} for a review) describes enhancement of a vector field by
advection and diffusion in a turbulent medium. Frozen-in advective transport of
a vector field by a chaotic flow generally enhances lengths along the field
lines exponentially in time. When that stretching effect prevails over resistive
diffusion, exponential growth of the magnetic energy occurs. The time constant
corresponds to the eddy turnover of the \emph{inner} turbulence scale. Thus, the
process runs to completion in a very short time, which only gets shorter as the
Reynolds number increases. The process yields a spectral distribution of
magnetic energy $P_B(k)$ that increases as $k^{3/2}$ up to the resistive cutoff
scale. That is, when the magnetic Prandtl number is high (as it is in the
interstellar medium), the early-stage magnetic energy is concentrated below the
viscous cutoff scale. In our simulations, the magnetic Prandtl number ranges from
$P_m \simeq 1 $ in the cold, dense star forming regions to $P_m \simeq 10^{25}$ near the 
outskirts of the galactic disk, while the typical $P_m$ for the ambient ISM is around
$10^{13}$. Once the small-scale magnetic energy density becomes
comparable to the small-scale turbulent kinetic energy density, the Lorentz
force begins to modify the turbulence in order to minimize further field line
stretching, and the process enters a non-linear stage. Local simulations of the
non-linear small-scale turbulent dynamo indicate that the magnetic energy
attains scale-by-scale equipartition with the kinetic energy after a few
turnovers of the largest eddies \citep{Beresnyak2012, Zrake2013,
  Schober2015}. In the case of a disk galaxy, the small-scale dynamo completion
time $\tau_{\rm dyn}$ is then $\sim h_{\rm disk} / v_{\rm eddy}$ where $h_{\rm
  disk}$ is the disk scale height and $v_{\rm eddy}$ is the RMS velocity of the
eddies at scale $h_{\rm disk}$.

The second panel of Figure \ref{fig:time_series} adds yet another piece of 
evidence for the small-scale dynamo explanation. Here we plot how the strength of
the poloidal component of the magnetic field as a fraction of the root-mean-square 
magnetic field strength varies with time. The poloidal component of the magnetic 
field oscillates in strength, and we do not observe the exponential growth that is
predicted for large scale dynamo action. We note that although the galactic magnetic
field is dominated by its toroidal component, that strength is set by the sum of many
small-scale, arbitrarily oriented toroidal field loops.

\subsection{Influence of the Star Formation Rate}
\label{sec:sfr}
The third panel of Figure \ref{fig:time_series} shows the star formation rate
as a function of time for each of the four models. It provides a valuable
complement to the first panel of that figure by informing us that the majority
of star formation took place within the first billion years of the galaxy's
formation. We note that the initial star formation rate of model g80LR is much
higher than those of the less-resolved models. The cause of this is purely
numerical. In g80LR, the highest resolved grid cell was unable to hold as much
mass as those in the less-resolved runs, which greatly stunted star particle
formation. To resolve this issue, we chose to make the minimum mass required for
star particles to be $2.5\times 10^5 M_{\odot}$. This correction led to g80LR
having roughly twice as many stars as the g320 and g160LR.  The star formation 
rate continues having an influence on the magnetic field strength and topology
even after the field saturates. Bursts of star formation contribute to the formation
of bumps in the total-to-injected magnetic energy ratio. High star formation is
also correlated with peak strengths of the poloidal magnetic field component. 
It is likely that increased turbulence from additional supernovae temporarily
augment the poloidal component of the magnetic field. 

Figure \ref{fig:B_rms} shows the density weighted root-mean-square value of the magnetic 
field as a function of time for all of the simulated model galaxies. All models
saturate around 6 $\mu G$ with the notable exception of models g80 and g80LR. We
believe that bursts of star formation after t = 2.0 Gyr in model g80LR (see third
panel of Figure \ref{fig:time_series}) and after t = 2.5 Gyr in model g80 provide the extra turbulence necessary to 
raise the magnetic field strength to roughly 15 $\mu G$.  We observe 
that models with the same minimum resolution share the same slope in the 
exponential growth of the magnetic field.  The magnetic field e-folding timescale ranges from 40 Myr to 215 Myr with different model resolutions. This corresponds to magnetic field growth rates of $\Gamma \simeq 1 \Omega - 6 \Omega$ assuming a galactic rotational period of 250 Myr. The magnetic field of g80 doesn't
reach $nG$ strengths until 1.2 Gyr, which is late compared to the other models. 
This is due to numerical dissipation around the small injection site and has no 
effect on the slope of the growth rate or the final magnetic field strength. 

Recent studies have shown that the Jeans
length scale needs to be resolved by at least 30 cells in order to capture the effect
of the turbulent dynamo \citep{Federrath:2011, Turk:2012}. The Jeans length is defined as

\begin{equation}\label{eqn:Jeans}
\lambda_J = \left( \frac{\pi c_s^2}{G \rho}\right)^{1/2},
\end{equation}

where $c_s$ is the sound speed of the gas, G is the gravitational constant, and $\rho$ is the density of the gas. 
From Figure \ref{fig:projections}, we estimate that the typical Jeans length in the galactic disk ranges from 
$\sim 5$ pc for the cold molecular gas to $\sim  3$ kpc for the warm ionized medium. Only our best-resolved 
models (g80 and g80LR) meet the 30 cell resolution criterion ($30 \times 80 \mathrm{pc} = 2.4$ kpc) in the warm 
phases of the ISM. Considering that the turbulent 
warm phase of the ISM covers a large fraction of the disk's volume, we interpret that the magnetic field growth in models 
g80 and g80LR is driven by the turbulent dynamo.


\subsection{Spectral Distribution of Magnetic and Kinetic Energy}
\label{sec:spectra}

\begin{figure}[t]
  \centering
  \leavevmode\epsfxsize=9.2cm\epsfbox{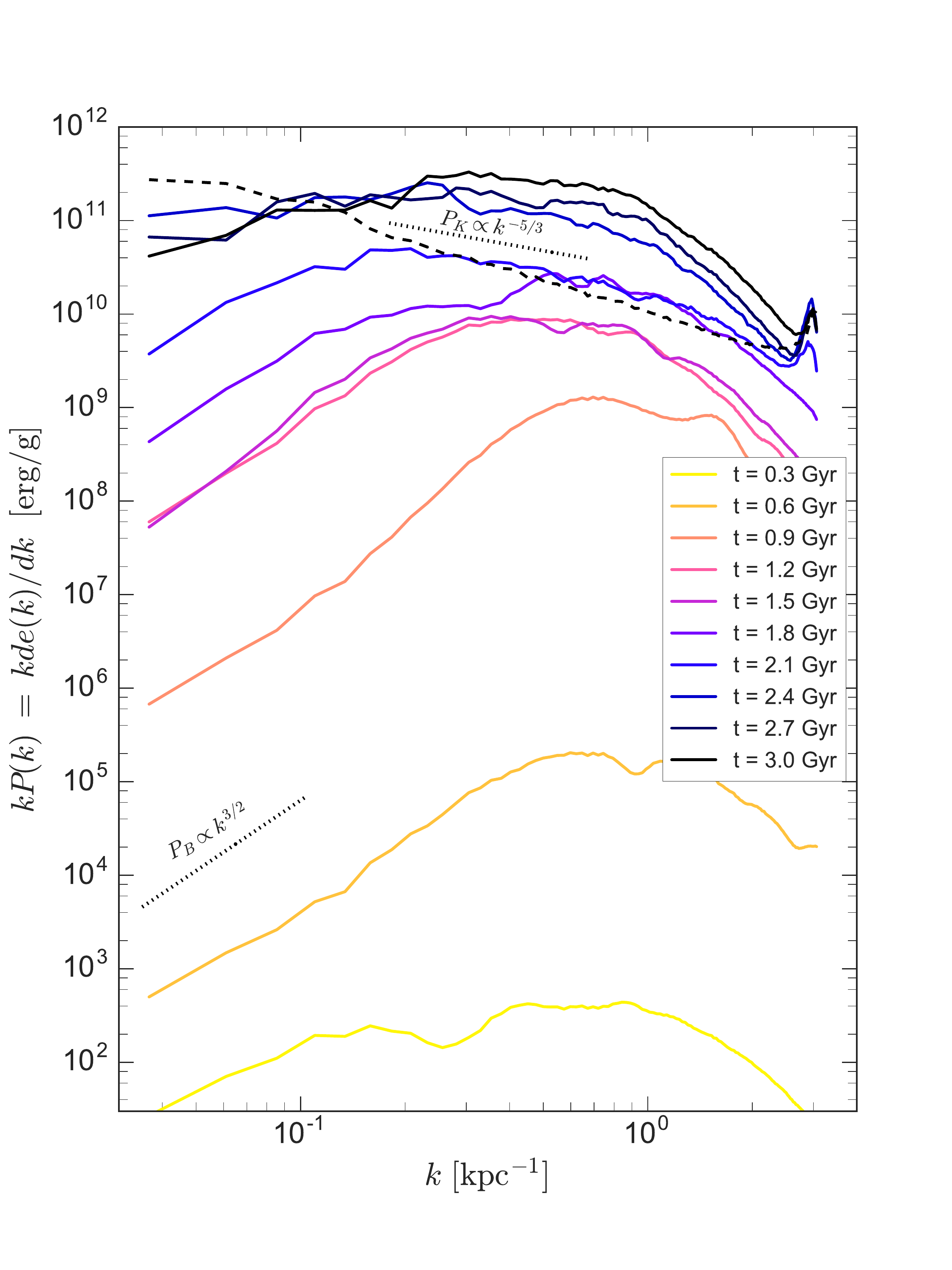}
  \caption{\footnotesize The time evolution of the magnetic energy power
    spectrum with evenly spaced time intervals from $t = \unit[300]{Myr}$ to $t =
    \unit[3]{Gyr}$ for galaxy model g80LR at time intervals of
    $\unit[300]{Myr}$. The dashed line shows the power spectrum of the kinetic
    energy at $t = \unit[3]{Gyr}$. The dotted lines provide visual references
    of the Kolmogorov and Kazantsev power laws. These labels correspond to power
    laws with those familiar indices, but the data used to create the curve is
    $k P_B(k)$, which is $\propto k^{a+1}$ where $a = 3/2$ or $-5/3$.
\label{fig:power_spectrum}
}
\end{figure}

In this section we present power spectra which indicate the scalewise
distribution of magnetic and kinetic energy. We briefly describe the procedure
by which the spectral diagnostics are obtained from simulation data, and then we
present results illustrating their evolution over time. In particular, we
discuss how the spectrum, and its time evolution provides evidence for the
operation of small-scale turbulent dynamo.

Power spectra are taken of the magnetic and kinetic energy per unit mass, which
are defined as
\begin{eqnarray} \label{eqn:power-spectra}
  P_B(k_i) &=& \frac{1}{\Delta k_i}\sum_{k_i < |\V q| < k_i + \Delta k_i} \V
  b_{\V q} \cdot \V b^*_{\V q} / 2, \\
  P_K(k_i) &=& \frac{1}{\Delta k_i}\sum_{k_i < |\V q| < k_i + \Delta k_i} \V
  v_{\V q} \cdot \V v^*_{\V q} / 2 , \nonumber
\end{eqnarray}
where $\V b_{\V q}$ and $\V v_{\V q}$ are the Fourier amplitudes of the
Alfv\'{e}n velocity $\V b = \V B / \sqrt{4 \pi \rho}$ and flow velocity $\V v$
respectively, at wavenumber $\V q$. The sum is over all wavenumbers whose
magnitude lies in the bin of width $\Delta k_i$ centered at $k_i$. Fourier
amplitudes are obtained using Fast Fourier Transform (FFT) routines provided by
the \emph{Numpy} module for Python. Since the simulation is performed using AMR,
solution data needs to be resampled onto a uniform rectilinear mesh before being
passed to the FFT. This latter stage is accomplished using the \emph{YT} package
to isolate a uniform data cube centered on the galactic disk. The data cube is
sampled at $512^3$ lattice points covering $(\unit[20.6]{kpc})^3$.

\begin{figure*}[t]
  \begin{minipage}{180mm}
    \centering
    \leavevmode\epsfxsize=8.9cm\epsfbox{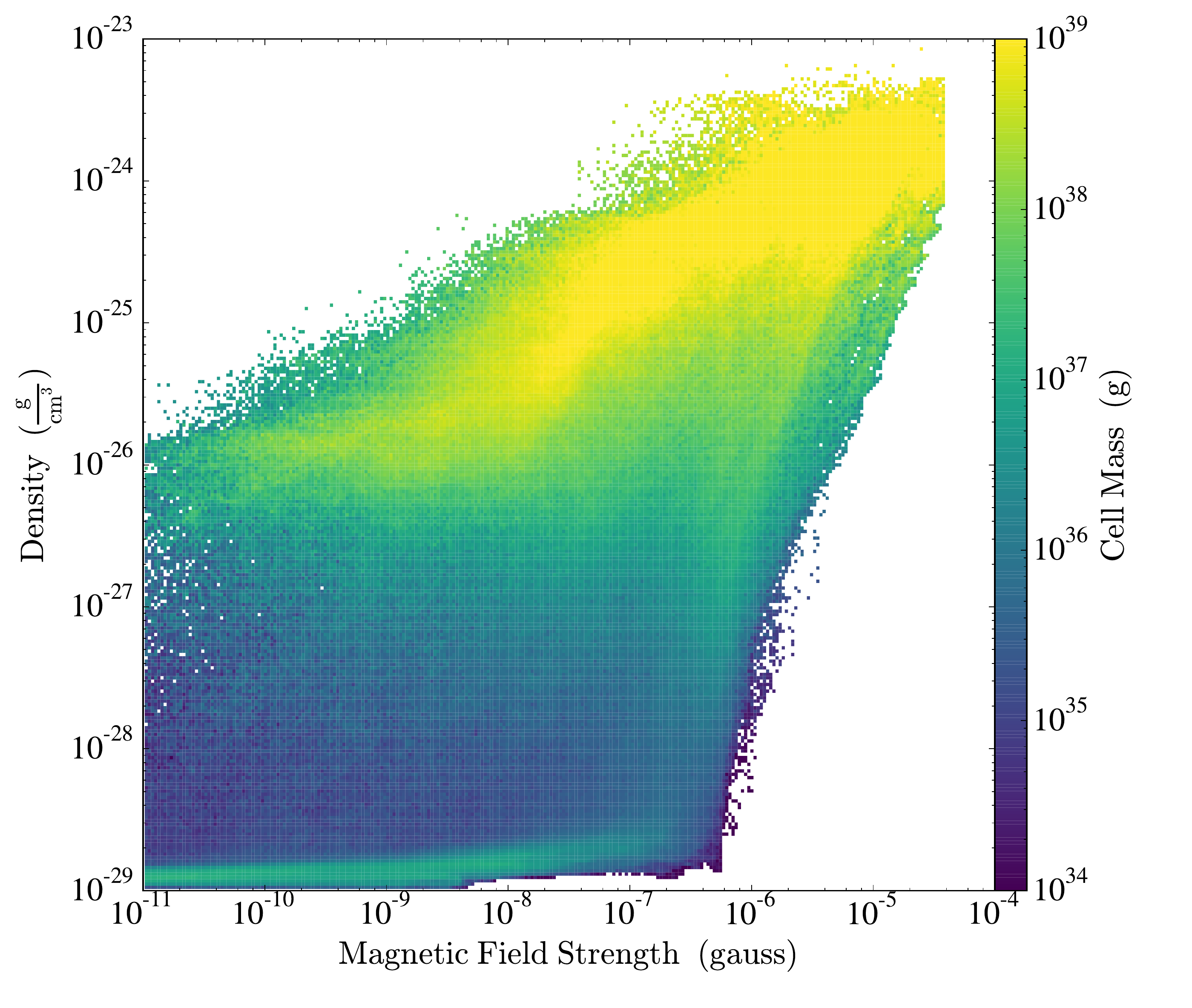}
    \leavevmode\epsfxsize=8.9cm\epsfbox{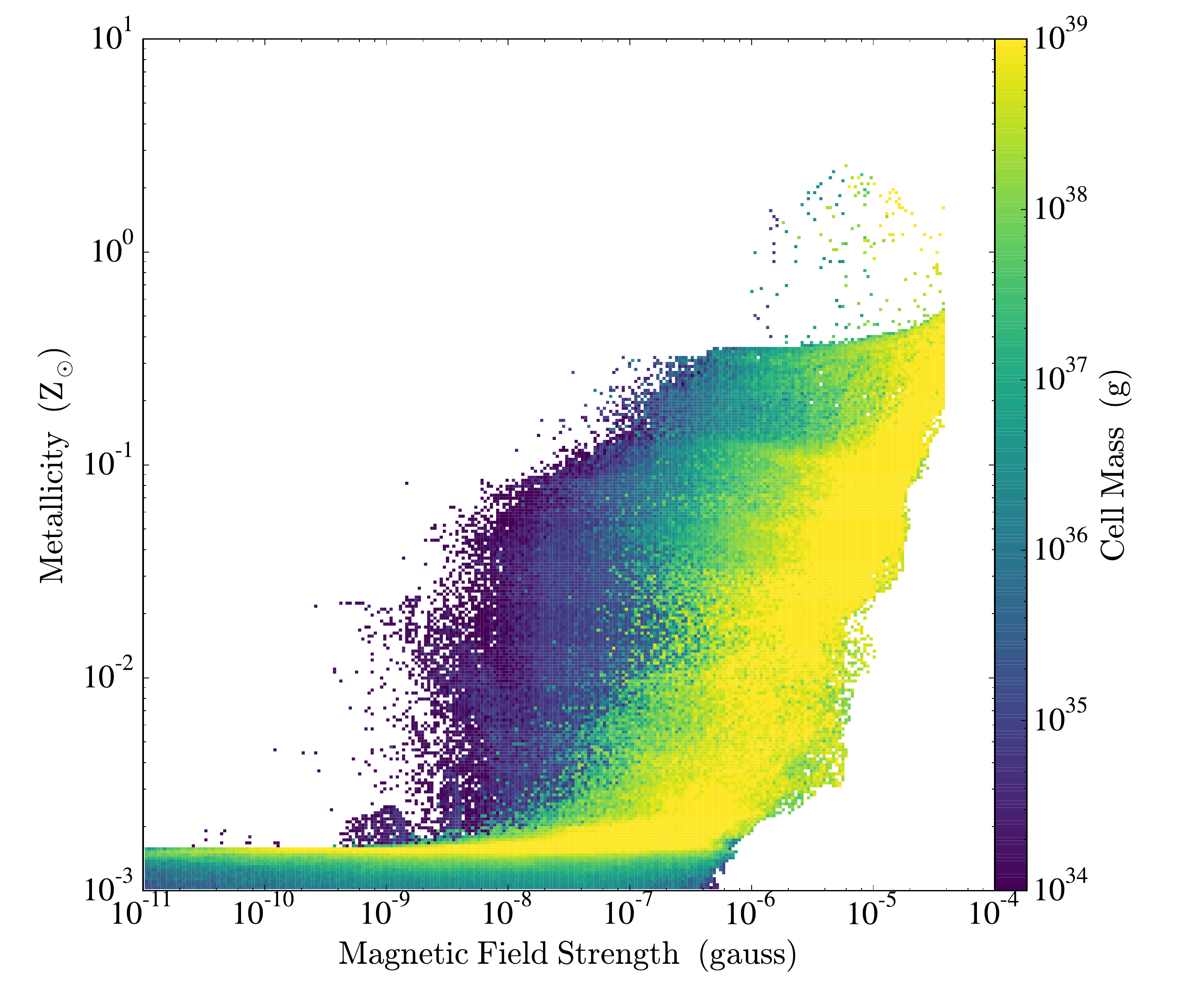}
    \caption{\footnotesize The distribution of cell mass as a function of 
    density and magnetic field strength {\it (left panel)} and metallicity 
    and magnetic field strength      {\it (right panel)}. The galaxy model
    pictured is g160LM after 3 Gyr of  evolution. The data was taken 
    from a sphere with a radius of 20 kpc about the center.
      \label{fig:phase-plots}
    }
  \end{minipage}
\end{figure*}

Figure \ref{fig:power_spectrum} shows the temporal evolution of the power
spectrum of the magnetic field energy as a function of the wavenumber k of the
model g80LR. The coloring of the solid lines denotes the time at which the
power spectrum was taken, starting with the yellow line at 0.3 Gyr, and ending
with the deep blue line at 3.0 Gyr in 0.3 Gyr increments. The dashed black line
represents the power spectrum of the kinetic energy of g80LR after saturation,
at 3.0 Gyr. The two dotted lines serve as visual aids for the Kazantsev ($P_B
\propto k^{3/2}$) and Kolmogorov ($P_K \propto k^{-5/3}$) power laws. General
consistency between the early time magnetic energy spectrum and the Kazantsev
spectrum suggests the magnetic field is being amplified by small-scale turbulent
dynamo action. The dashed line in Figure \ref{fig:power_spectrum} shows the
kinetic energy spectrum, which is consistent with the Kolmogorov spectrum. 
The peak of the magnetic spectrum at saturation ($k_* \simeq 4 \times 10^{-2}$ kpc$^{-1}$ 
lies at approximately 10 times the forcing scale, $k_L$. Also
note that the late-time magnetic energy spectrum indicates that the magnetic
field is marginally dominant at the small scales, and marginally sub-dominant at
the large scales. These features are all consistent with the completion of
non-linear small-scale turbulent dynamo in the saturation regime for
high magnetic Prandtl numbers \citep{Schekochihin:2004, Schober:2012, Schober:2015}.
However, the geometry of our thin-disk galaxies poses a challenge in finding
truly isotropic properties of magnetic and kinetic turbulence. The growth of the
magnetic power spectrum could also be influenced by an efficient small-scale
injection of the magnetic field, which is transported to larger scales through
shear and rotation. Without performing control runs, our interpretation of the
role of a turbulent small-scale dynamo remains speculative.

\subsection{Metallicity-Magnetic Field Correlations}
\label{sec:metallicity}

The left side of Figure \ref{fig:phase-plots} shows a 2D histogram of samples of
the density and magnetic field strength throughout the disk. Each data point is also given a
color corresponding to the cell mass at the sample location. In
the right panel, the same samples are scattered according to their field
strength and metallicity value, and colored according to the mass in a given cell. Data
is shown from the low metallicity model, g160LM, after it was evolved for 1
Gyr. We choose g160LM because of its low initial metallicity, which makes it an 
excellent candidate to trace the contribution of metals from supernova events. 
At 1 Gyr, the magnetic field strength is attaining its saturated value
(the time series data for g160LM is essentially the same as for g160, shown in
Figure \ref{fig:time_series}). The left panel reveals that the magnetic field is
strongest in the dense regions of the galaxy, where star formation is
prevalent. Meanwhile, the field strength and metallicity are weakest in the
less-dense regions of the galaxy, which are found in the outskirts of
the galactic disk and in the outer halo. From the right panel, we see that the
magnetic field strength correlates with higher metallicity values. Though we
already know supernova ejecta to be our sources of metals, we note that the
magnetic field and high metallicity \emph{remain spatially coupled}, even after
the supernova events have ended. We note that
this effect depends on the mixing efficiency of the ISM which operates on 
sub-grid scales and is therefore difficult to simulate effectively. Nevertheless, this is a distinct prediction made by the SN
seed field / SN-driven dynamo scenario we have simulated for this study; if the
seed field had been supplied uniformly in the initial condition, no magnetic
field strength-metallicity correlation would be anticipated.

\subsection{All-sky Rotation Measure Map}
\label{sec:rotation-measures}

Magnetic fields in simulated galaxies such as the one reported in this article
could be compared with actual observed data.  The most detailed such
galaxy-scale magnetic field observation is of our Milky Way galaxy.  In this
section, we compare our simulation results with Milky Way's magnetic fields by
constructing an all-sky map of Faraday rotation measures.

When electromagnetic waves travel through a magnetized interstellar medium,
their planes of polarization rotate, and the amount of rotation is dependent
upon the intervening magnetic field.  Rotation measure (RM), defined below, is
proportional to the magnetic field strength parallel to the line of sight, and
the density of thermal electrons.
\begin{align}
  {\rm RM} = \frac{e^3}{2\pi m_e^2 c^4} \int^L_0 n_e(s) B_{\parallel}(s)
  ds \label{Eq: RM}
\end{align}
Here $n_e$ is the thermal electron number density, $m_e$ is the mass of an
electron, $B_{\parallel}$ is the magnetic field parallel to the line of sight,
and $s$ is the line segment along the line of sight.  Below we start by
describing how this RM is evaluated in each lines of sight and how the all-sky
RM map is constructed for our simulated galaxies.

\begin{figure}[t]
  \includegraphics[width = 8.8 cm,clip=true,angle=0]{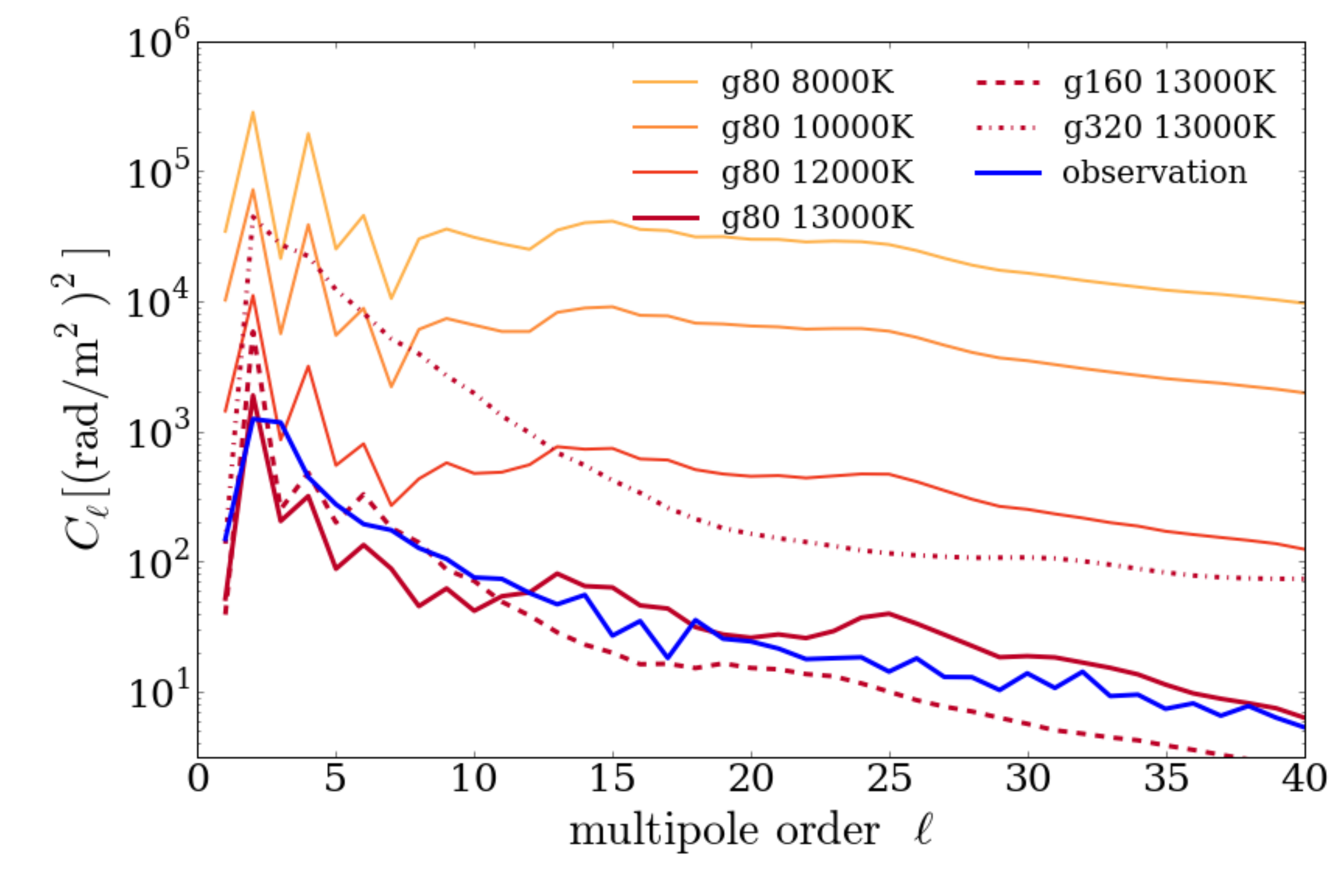}
  \caption{Angular power spectra of the simulated all-sky Faraday rotation
    measure (RM) map.  A series of yellow to red solid lines are for the g80LR run
    with varying ionization threshold temperatures $T_{\rm thres}=\unit[8000]{K}
    - \unit[13,000]{K}$ at $\sim \unit[2.4]{Gyr}$.  The blue solid line is for
    the Milky Way observation \citep{Oppermann:2012aa} whose corresponding
    all-sky map is reproduced in Figure \ref{fig:milky way rm}.  The dashed and
    dot-dashed lines are for g160LR and g320 runs, respectively, with $T_{\rm
      thres}=\unit[13,000]{K}$. The all-sky maps corresponding to g320, g160LR,
    g80LR runs with $T_{\rm thres}=\unit[13,000]{K}$ are Figures \ref{fig:g320},
    \ref{fig:g160}, and \ref{fig:g80} respectively.}
  \label{fig: power spectra}
\end{figure}

In order to calculate RM, one first needs to estimate the number density of
thermal electrons.  Since we do not explicitly trace electron species in our
simulations but only follow total gas density $\rho_{\rm gas}$, we choose to use
the specific thermal energy of gas (per unit mass), $E_{\rm th}$, to estimate
the ionization fraction $\chi(E_{\rm th})$ of hydrogen atoms, and then the
thermal electron density $n_e$ in each cell as
\begin{align}
  n_e = \frac{\rho_{\rm gas}\chi(E_{\rm th})}{m_{\rm H}}
  \label{Wq: ne(chi)}
\end{align}
where $m_{\rm H}$ is the mass of a hydrogen atom.

If we assume that hydrogen becomes instantaneously fully ionized at temperature
$T=T_{\rm thres}$, one can show that the ionization fraction is related to
$E_{\rm th}$ as
\begin{align}
  \chi(\mu_{\rm eff}(E_{\rm th})) =\left\{
  \begin{array}{lr}
    0 & \text{for } \mu_{\rm eff} \geq \mu_{\rm max}\\ \frac{4}{3 \mu_{\rm eff}}
    - \frac{13}{12} \;\;\;\;\;\; & \text{for } \mu_{\rm min} \leq \mu_{\rm eff}
    \leq \mu_{\rm max}\\ 1 & \text{for } \mu_{\rm eff} \leq \mu_{\rm min}
  \end{array}
  \right.
  \label{Eq: chi(mu)}
\end{align}
with an {\it effective} mean molecular weight function defined as
\begin{align}
  \mu_{\rm eff}(E_{\rm th}) = \frac{3}{2} \frac{ N_{\rm A} k_{\rm B} T_{\rm
      thres}}{E_{\rm th}},
  \label{Eq: mu(E)}
\end{align}
where $N_{\rm A}$ is the Avogadro number, and $k_{\rm B}$ the Boltzmann
constant.  Here, if all hydrogen atoms are neutral, the mean molecular weight is
$\mu_{\rm max} \simeq 1/(0.25/4 + 0.75/1) = 1.23$, whereas if all hydrogen atoms
are ionized, the mean molecular weight is $\mu_{\rm min} \simeq 1/(0.25/4 +
0.75/1\times2) = 0.64$.  Note that the only free parameter in our formulation
Eq.(\ref{Eq: chi(mu)}) is $T_{\rm thres}$, which should be $\sim 10^4\,{\rm K}$.

\begin{figure}[t]
  \centering \includegraphics[width = 8.5
    cm,clip=true,angle=0]{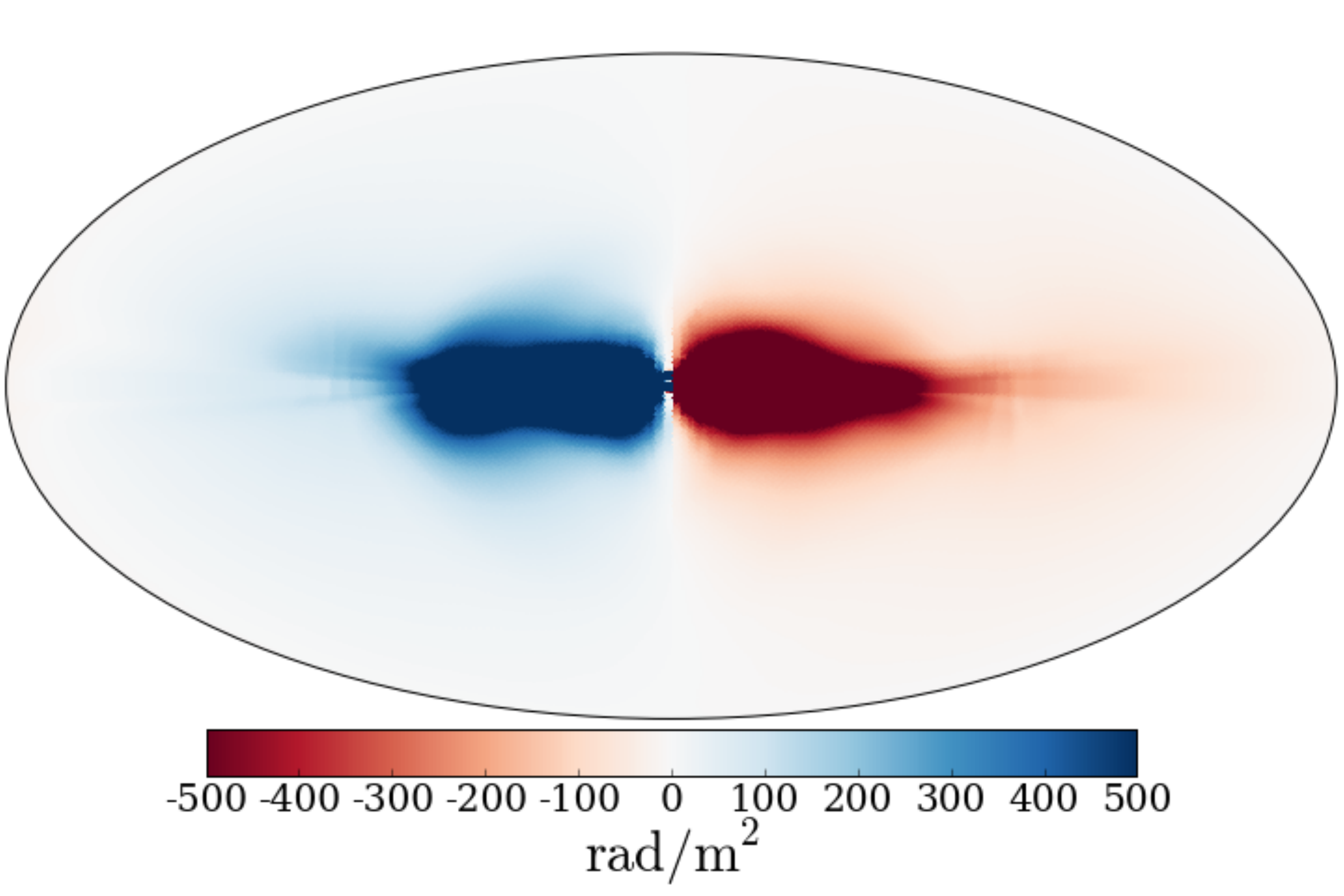}
  \caption{All-sky RM map for the g320 run at $\sim 2.4 \,{\rm Gyr}$ with
    $T_{\rm thres}=\unit[13,000]{K}$ and an origin 8 kpc away from the center of
    the galaxy. The angular power spectrum of this map is shown in Figure
    \ref{fig: power spectra}. }
  \label{fig:g320}
\end{figure}

\begin{figure}[t]
  \centering \includegraphics[width = 8.5
    cm,clip=true,angle=0]{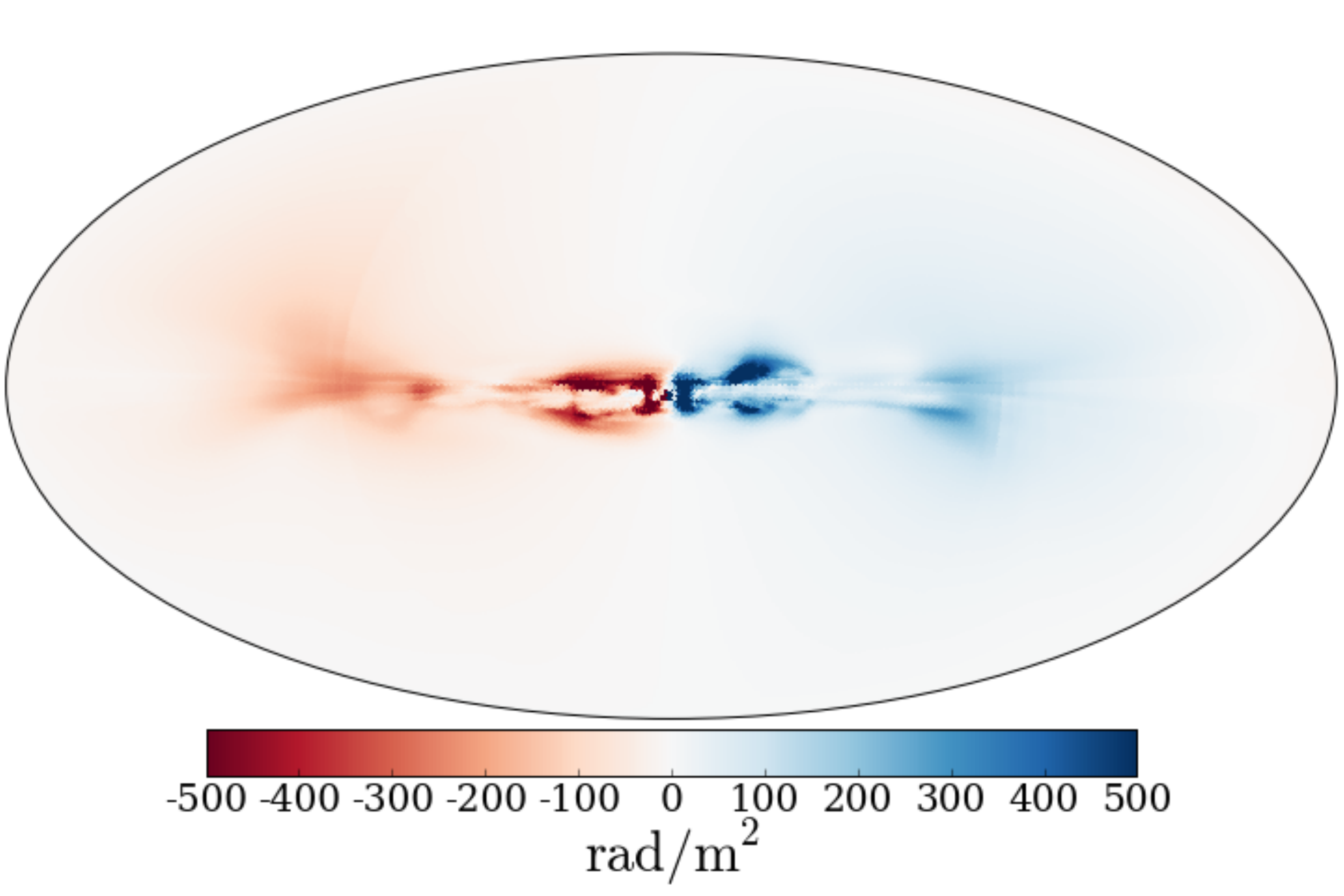}
  \caption{All-sky RM map for the g160LR run at $\sim 2.4 \,{\rm Gyr}$ with
    $T_{\rm thres}=\unit[13,000]{K}$ and an origin 8 kpc away from the center of
    the galaxy. The angular power spectrum of this map is shown in Figure
    \ref{fig: power spectra}. }
  \label{fig:g160}
\end{figure}

\begin{figure}[t]
  \centering \includegraphics[width = 8.5
    cm,clip=true,angle=0]{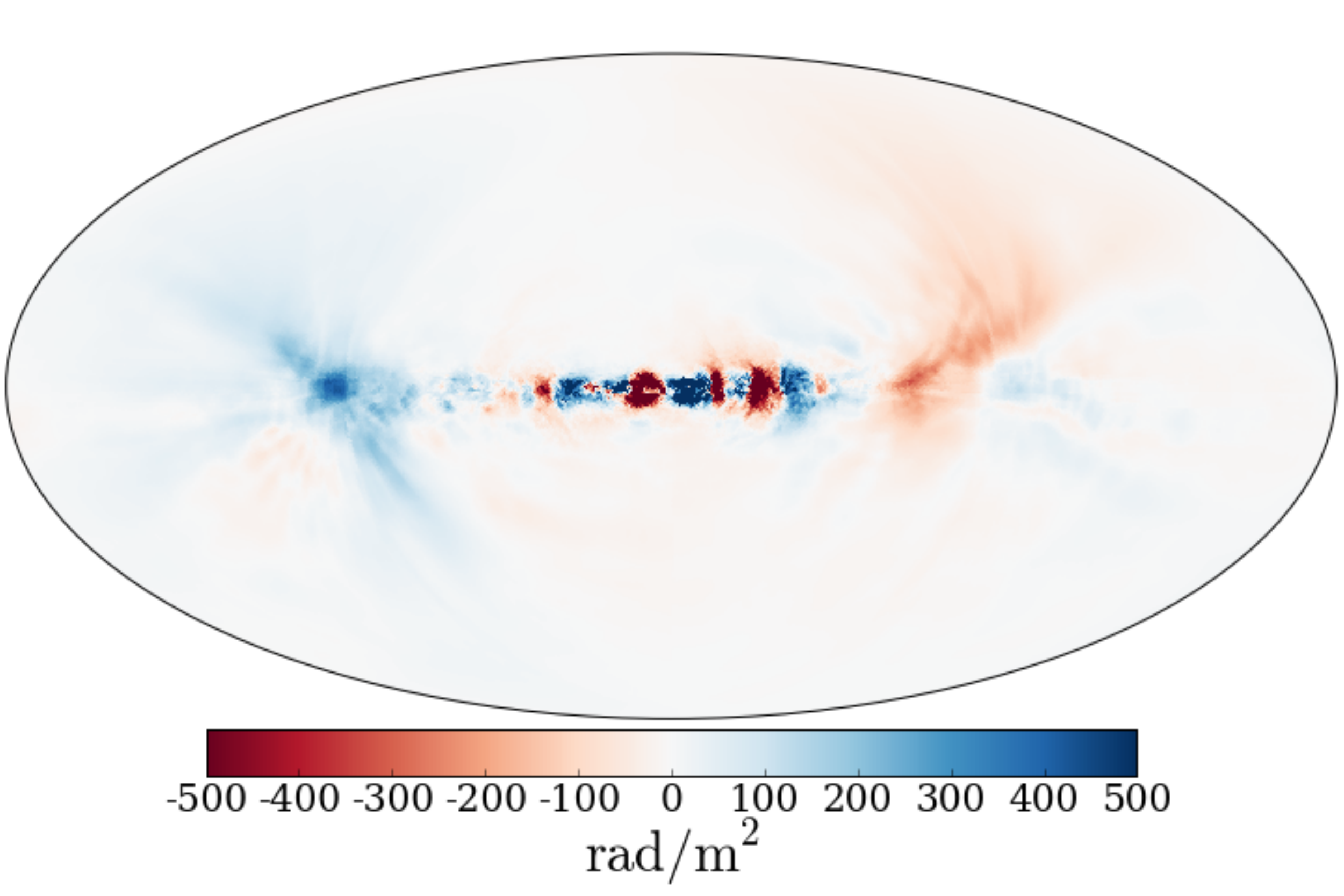}
  \caption{All-sky RM map for the g80LR run at $\sim 2.4 \,{\rm Gyr}$ with $T_{\rm
      thres}=\unit[13,000]{K}$ and an origin 8 kpc away from the center of the
    galaxy. The angular power spectrum of this map is shown in Figure \ref{fig:
      power spectra}. }
  \label{fig:g80}
\end{figure}

With Eqs.(\ref{Wq: ne(chi)}) to (\ref{Eq: mu(E)}), one can evaluate the thermal
electron density $n_e$ from the specific thermal energy $E_{\rm th}$.  We then
integrate $n_e(s) B_{\parallel}(s)$ along the line of sight as in Eq.(\ref{Eq:
  RM}) to acquire the rotation measure, then to produce an all-sky map like
Figures \ref{fig:g320} to \ref{fig:g80}.  But first, in Figure \ref{fig: power
  spectra}, we plot the angular power spectra of all-sky RM maps for the g80LR run
with 4 different ionization threshold temperatures ranging $T_{\rm
  thres}=8000\,{\rm K}$ to $\unit[13,000]{K}$ (yellow to red solid lines), in
order to demonstrate how sensitive an RM map is to our choice of $T_{\rm
  thres}$. This figure is for $t \sim \unit[2.4]{Gyr}$, well after galactic
magnetic fields have reached nominal equilibrium values (see Figure
\ref{fig:time_series}).  RM is proportional to thermal electron density, and we
find that RM is highly sensitive to our understanding of how gas is ionized
which is only crudely parametrized by $T_{\rm thres}$ in our model, Eq.(\ref{Eq:
  chi(mu)}).  This illustrates the need to carefully characterize the ionization
states of intervening medium to correctly understand the galactic magnetic
field, which however is beyond the scope of this study.  For the purpose of the
current article, we opt to simply utilize the fact that $T_{\rm
  thres}=\unit[13,000]{K}$ best matches the observed Milky Way power spectrum
assembled by \citet{Oppermann:2012aa}, shown here as a blue solid line.
Therefore, we adopt $T_{\rm thres}=\unit[13,000]{K}$ for subsequent figures and
discussion hereafter.  Also in Figure \ref{fig: power spectra}, the power
spectra for g160LR and g320 runs are plotted for comparison with $T_{\rm
  thres}=\unit[13,000]{K}$.

\begin{figure}[t]
  \centering \includegraphics[width = 8.5
    cm,clip=true,angle=0]{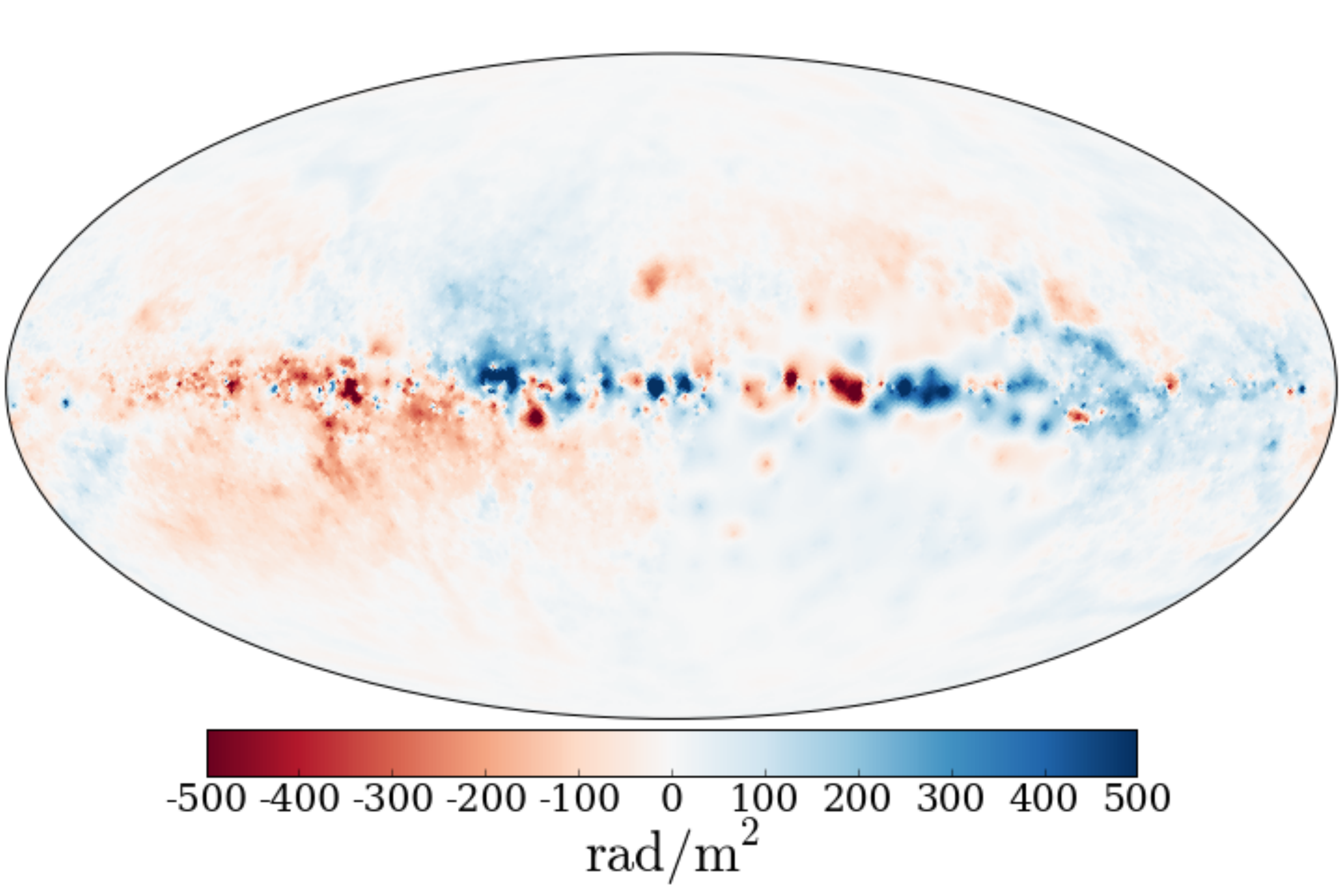}
  \caption{All-sky RM map for the Milky Way galaxy, recreated from
    \citet{Oppermann:2012aa}. The angular power spectrum of this map is shown in
    Figure \ref{fig: power spectra}. }
  \label{fig:milky way rm}
\end{figure}

To construct simulated all-sky RM maps that mimic \citet{Oppermann:2012aa}, we
position an observer at an approximate location of the Sun in our Milky Way: on
the galactic disk plane about 8 kpc away from the Galactic Center.  For this
purpose, we first identify the location of highest density peak, $(x_*,\, y_*,\,
z_*)$, and define $z = z_*$ as the galactic disk plane.  The origin (center) of
the RM maps is then set on this plane, approximately 8 kpc from the galactic
center in $x$-axis.\footnote{This translates into $(0.494, \,0.5,\, z_*)$ in [0,
    1] coordinates.  The galactic center is initially placed at $(0.5,\, 0.5,\,
  0.5)$, but as the simulation progresses it may shift to another location with
  minor offsets varying in different runs.  This is why an extra step is needed
  to first accurately determine the galactic disk plane.}  By employing the {\sc
  Healpix} pixelisation scheme \citep{Gorski:2005}, Figures \ref{fig:g320} to
\ref{fig:g80} display the resulting all-sky RM maps at $\sim \unit[2.4]{Gyr}$
for g320, g160LR and g80LR runs, respectively, with the ionization threshold
temperature $T_{\rm thres}=\unit[13,000]{K}$.  All these maps have the same
angular resolution (i.e.  {\sc Healpix} parameter $N_{\rm side} = 128$) that
matches the observed Milky Way map in Figure \ref{fig:milky way rm}.  This
Figure \ref{fig:milky way rm} is recreated from \citet{Oppermann:2012aa} for
comparison, using the raw data publicly available at {\tt
  http://wwwmpa.mpa-garching.mpg.de/ift/faraday}.

Readers should be cautioned that the our RM maps and the corresponding power
spectra are made for one particular snapshot ($\sim 2.4 \,{\rm Gyr}$) of a
greatly idealized disk galaxy, centered at an almost arbitrary location ($\sim$
8 kpc from the galactic center in $x$-axis).  Yet, several similarities are
already noticeable between simulated and observed maps, such as the prominent
magnetic field structure in disks, the existence of coherent large-scale field
directions, and numerous smaller-scale structures possibly associated with
turbulent small-scale eddies.  The resemblance is the strongest for the g80LR run
that has the smallest spatial resolution in our simulation suite.  In
particular, for the g80LR run, the spectral distribution of power in the multipole
$l$-space is remarkably similar to the observed RM map, --- albeit with a tuned
$T_{\rm thres}$ value ---, as can be seen in the angular power spectra of Figure
\ref{fig: power spectra}. The model deviates from observations in the outer
galaxy, where there is a lack of RM signal, both in strength and structure. This can be
the result of several different factors. The first possibility is that the resolution of the grid
decreases away from the galactic disk, which would wash out small-scale structure.  The second is that the strength of the 
magnetic field outside of the galaxy is very weak along the plane of the disk because 
the magnetic field is ejected primarily through jets along the axis normal to the face of the disk.  
Since our snapshot is taken after 2.4 Gyr, there may not have been enough time for the 
surrounding medium to gain enough magnetic strength. Finally, our approximation of the ionization
fraction neglects photoionization processes which would lead us to underpredict the RM signal in hot,
low-density regions.
Despite a highly idealized set-up, we conclude that
our simulations reproduce an RM map and power spectrum that are close in shape
to Milky Way's --- especially for a run with the finest resolution available.


\section{Discussion} \label{sec:discussion}

\subsection{Comparison With Other Computational Approaches}

Our approach is similar in certain ways to that taken by other groups.
\cite{Hanasz2009} have also performed global disk galaxy simulations with
magnetic fields introduced by supernovae. Whereas we used a toroidal magnetic
configuration for the magnetic source term, they used randomly oriented magnetic
dipoles. Their focus was on the role of cosmic rays in driving the galactic
dynamo, so instead of introducing thermal energy as we did in our feedback
prescription, they included a cosmic ray fluid that is sourced in the vicinity
of supernova explosions. In their approach, only 10\% of supernova events
included a magnetic field. The same approach was adopted in
\cite{Kulpa-Dybe2011, Kulpa-Dybe2015} to simulate dynamo activity in barred
spiral galaxies, and by \cite{Siejkowski2014} to study dwarf galaxies.
\cite{Beck:2013aa} also utilized a magnetic dipole configuration as supernova
source terms. Their simulations were carried out using an MHD version of the
smoothed particle hydrodynamics (SPH) code \emph{GADGET}, and were focused on
magnetic evolution in a protogalactic halo instead of a disk
galaxy. \cite{Wang2009, Rieder2016} carried out global disk galaxy simulations
of a dynamo driven by hydrodynamic stellar feedback. In those studies a
large-scale magnetic field was imposed on the initial model, rather than being
seeded by the supernovae themselves. All studies appear to be in agreement
regarding the saturation time-scale for turbulent dynamo.

\subsection{Large Scale Magnetic Field}

Throughout the course of our simulations, the magnetic field develops long-range
structure. This behavior is somewhat surprising, given that small-scale
turbulent dynamo, which we believe to be the driving mechanism, is not known to
excite magnetic field modes of longer wavelength than that of the
energy-containing turbulent eddies ($\lesssim h_{\rm disk}$). This could
suggests that something akin to a mean-field dynamo (e.g. $\alpha-\Omega$) may
be in effect. To determine whether that is the case, one could, in principle,
measure the velocity-vorticity correlation from the simulation data to see if it
has sufficient amplitude to explain the strength of the long wavelength magnetic
field. Such a diagnostic could be a computational test of the arguments
developed by \cite{Kulsrud1992}, who pointed out that mean-field dynamo (which
is a linear theory) may be quenched by back-reaction of the field, even at
amplitudes well below equipartition.

Nevertheless, there are other mechanisms that may be responsible for the
large-scale magnetic field structure seen in our simulations. As we saw in
Section \S \ref{sec:metallicity}, the field strength is correlated with gas
density. This suggests that the field enhancements could simply be tied to the
density structure of the disk, which itself contains a long-wavelength component
arising from self-gravitating effects. Another possibility is that the long-range
field is simply the result of relaxation in regions where star formation is less
active. There are at least two facts to support this view. First, it has been
established observationally that the magnetic field of star-forming galaxies is
generally stronger and more scrambled than in late-type galaxies
\citep{Chyzy2007}. Secondly, the condition for magnetic fields to exhibit
self-assembly (inverse energy transfer) during turbulent relaxation turns out to
be weaker than was previously believed. Although the presence of net magnetic
helicity is known to be sufficient for inverse transfer to occur during
relaxation \citep{Frisch1975, Alexakis2006}, it is now understood that is not a
necessary condition; magnetic coherency increases due to statistical merging of
locally relaxed magnetic structures even in non-helical field configurations
\citep{Zrake2014, Brandenburg2015}. Thus, the development of large-scale
structure, at least within regions of the disk where turbulence is not actively
driven, should be anticipated even if the net helicity is negligible; it is
unnecessary to invoke possible mechanisms by which magnetic helicity may be
expelled from the disk \citep[e.g.][]{Sur2007}.

\subsection{Appearance of the Earliest Galactic Magnetic Fields}

In the scenario we have simulated here, magnetic fields are first introduced
alongside the first supernovae, which in our simulation occur in a fully formed,
yet unmagnetized Milky Way type disk galaxy. This is not historically accurate,
since the first stars and supernovae occurred at a much earlier epoch. Thus, our
study should be interpreted as a proof of principle, that active star formation
maintains galactic magnetic fields around equipartition levels. Since that star
formation begins at an epoch preceding the disk galaxy phase we simulated here,
a realistic narrative is that the protogalaxies that accreted to form the Milky
Way were probably magnetized according to their level of star formation, and so
the galactic disk inherited its first magnetic field from its progenitors.  If
agitation by supernova feedback were to terminate, then the magnetic field would
relax and dissipate over the same time scale (the disk scale-height eddy
turnover time) as it was built up. So, the strength and morphology of the
magnetic field we observe in the evolved disk galaxy is arguably insensitive to
initial conditions; had we begun the simulation with a disk magnetized at
equipartition, supernova feedback would be expected to maintain it.

\subsection{Role of Cosmic Rays}

Our simulations use a stellar feedback prescription in which thermal energy and
(small amounts of) magnetic energy are supplied by supernova explosions, but
cosmic rays (CR's) are not included. Interest in the role of CR's in driving a
galactic dynamo goes back to \cite{Parker1992}, who pointed out that their
buoyancy can lead to stretching of the field which enhances its long-wavelength
component. Several numerical studies have now been conducted \citep{Hanasz2009, 
Siejkowski2014, Kulpa-Dybe2011, Kulpa-Dybe2015} that include CR
effects coupled with MHD cosmological simulation. Overall, the results are
surprisingly similar to our own; magnetic fields reach equipartition levels over
$\unit[]{Gyr}$-time scales. This raises the question of the relative importance
of the two primary types of energy supplied by supernovae --- turbulence driven
by the thermal expansion of the remnants themselves, and energy released as CR's
--- since either by itself appears to be sufficient.

\section{Conclusion} \label{sec:conclusion}

The existence of galactic magnetic fields is a significant puzzle in
astrophysics and cosmology. In this paper, we have shown how those fields could
have reached their present-day strength and morphology by stellar feedback
processes alone. Weakly magnetized plasma is introduced to the galactic medium
by stellar winds and supernova explosions, and the magnetic field is
subsequently amplified by turbulence, driven by those same outflows. We carried
out simulations of an isolated Milky Way type disk galaxy, that
was initiated in an unmagnetized state. The galaxy was then evolved through an
epoch of star formation and associated magnetohydrodynamic (MHD) feedback. Our
simulations were continued through $\unit[3]{Gyr}$, at which time $\unit[]{\mu
  G}$-level magnetic fields were seen throughout much of the disk. Analysis of
the time series and power spectrum of the magnetic field supports a picture in
which small-scale turbulent dynamo, with turbulence sustained by supernova
feedback, is the primary means by which the field reaches dynamically relevant
strengths. The presence of differential rotation keeps the field marginally
dominated by its toroidal component. In the scenario we have studied here,
magnetic fields are injected in metal-rich supernova ejecta. Our simulations
predict that a correlation between magnetization and metallicity should persist
throughout galaxy's evolution, and we argue that correlation is a distinct
signature of a galactic dynamo seeded by supernova feedback. Finally, we presented
synthetic all-sky maps of the Faraday rotation measure, as it would appear from
Earth's location relative to the galactic center. The angular power spectrum is
found to match well to observations when the plasma ionization threshold
temperature is chosen to be $\unit[13,000]{K}$.

In this work, we have also described a novel computational technique for
simulating MHD supernova feedback processes. The simulation initial data we used
is publicly available as part of the AGORA project. The initial model was
evolved using the MHD version of the \emph{ENZO} cosmological framework. We
used the stellar formation algorithms that exist in the public version of that
code, but modified the feedback prescription to include a magnetic field source
term. The source term is chosen to inject a small fraction of the supernova
energy as toroidal magnetic field, oriented relative to a randomly chosen
vertical axis for each supernova event. The amplification and distribution of
magnetic field in the evolved disk galaxies was found to be insensitive to the
chosen injection radius or the initial metallicity of the ambient medium. 
Instead, the saturated field strength appears to be controlled by
the level of turbulence sustained by stellar feedback. We also examined the
influence of the magnetic field injection footprint size. Since resolution fine
enough to resolve $\sim \unit[]{pc}$-scale supernova remnants remains
computationally prohibitive, we ran a family of models with increasing
resolution, and decreasing footprint size, between $\unit[320]{pc}$ and
$\unit[80]{pc}$. The footprint size appears not to influence the outcome,
suggesting that our conclusions would not change significantly if the injection
scale was somehow reduced to astrophysically realistic values.

\subsection{Acknowledgements}
The authors thank Romain Teyssier, Michael Rieder, Klaus Dolag, and the anonymous referee
for their insightful comments. I.B. would like to thank Sam Skillman for his time 
and help with \emph{ENZO}.This work was performed in part under DOE 
Contract DE-AC02-76SF00515 and support by the Kavli foundation. 
The simulations were run on the Bullet cluster at the SLAC National
Accelerator Laboratory and the Sherlock cluster at Stanford University.




\bibliography{MagneticFieldReferences,ZrakeReferences}
\bibliographystyle{apj}

\end{document}